\documentclass[preprint2]{aastex}

\newcommand\plotfour[4]{\centering \leavevmode
 \columnwidth=1.99\columnwidth 
 \includegraphics[width={0.24\columnwidth}]{#1}%
 \includegraphics[width={0.24\columnwidth}]{#2}%
 \includegraphics[width={0.24\columnwidth}]{#3}%
 \includegraphics[width={0.24\columnwidth}]{#4}%
}%

\newcommand\plotwide[1]{\centering \leavevmode
 \columnwidth=1.99\columnwidth 
 \includegraphics[width={\columnwidth}]{#1}
}

\begin{document}

\title{Simulation of radio plasma in clusters of galaxies}

\author{M.Br\"{u}ggen\altaffilmark{1,2,3}, C.R.Kaiser\altaffilmark{4}, E.Churazov\altaffilmark{1,5}, T.A.En{\ss}lin\altaffilmark{1}} 

\altaffiltext{1}{MPI f\"{u}r Astrophysik, Karl-Schwarzschild-Strasse 1, 85740
Garching, Germany}
\altaffiltext{2}{International University Bremen, Campus Ring 1, 28759 Bremen, Germany}
\altaffiltext{3}{Churchill College, Storey's Way, Cambridge CB3 0DS, UK}
\altaffiltext{4}{Department of Physics \& Astronomy, University of Southampton,
University Road, Southampton SO17 1BJ, UK} 
\altaffiltext{5}{Space Research Institute (IKI), Profsouznaya 84/32,
Moscow 117810, Russia}

\begin{abstract}
We present three-dimensional hydrodynamical simulations of buoyant gas
in a typical cluster environment. The hot matter was injected
continuously into a small region off-set from the cluster centre. In
agreement with previous analytic estimates we found that the bubbles
evolve very differently depending on their luminosity. Using tracer
particles we computed radio maps of the bubbles based on different
assumptions about the magnetic field. In the radio band the bubbles
closely resemble FRI sources. For the bubbles to be detectable for
long enough to account for FRI sources, we found that reacceleration
has to take place. The bubbles are generally difficult to detect,
both, in the radio and in the X-ray band. Thus it is possible to hide
a significant amount of energy in the form of bubbles in
clusters. Finally, we compute the efficiency of the bubbles to stir
the ICM and find that recurrent low-power sources may be fairly
effective in mixing the inner cluster region.

\end{abstract}


\label{firstpage}

\sloppypar

\section{Introduction}

The X-ray surface brightness of many clusters of galaxies shows a
strong central peak which is generally interpreted as the signature of
a cooling flow (Cowie \& Binney 1977, Fabian \& Nulsen 1977, Sarazin
1988, Fabian 1994). However, the simple cooling flow model conflicts
with a growing number of observations that show that gas with a
temperature below 1 keV is significantly less abundant than
predicted. Searches for cold gas in the infrared and radio have have
detected far less mass than detected. Therewhile, star formation can
only account for a few percent of the predicted cooling rate.

Many clusters of galaxies host a radio source at their centres. These
radio sources emit large quantities of hot relativistic plasma into
the intracluster medium (ICM). However, the problem of estimating the
effect of the radio plasma on the ICM is substantially more intricate
and difficult than the simple cooling flow model (Binney 2001). The
radio source destroys the assumed spherical symmetry of the cooling
flow. This makes the geometry more complex which is aggravated by the
fact that magnetic fields with complex morphologies are likely to play
an important role. Moreover, the radio source introduces dynamical
motions into the medium pushing it out of hydrostatic equilibrium. To
make matters worse, the source may even be time-dependent as many
radio sources are believed to go through active and less active
cycles.  It is therefore difficult to come up with robust
predictions. Nevertheless, the problem is important and, with the
advent of more powerful X-ray telescopes, timely enough to explore it
even with simplified models. Meanwhile, evidence for bubbles in the
ICM is accumulating, especially from the X-ray telescope Chandra
(Fabian et al. 2000, Mazzotta et al. 2001).

A hydrodynamic simulation of the full problem is currently beyond our
possibilities as the scales involved vary enormously: The velocities
range from around 10 km/s in the outer cooling flow region to close
to the speed of light in the radio jet. The densities and energy
densities likewise vary by many orders of magnitude. Thus one may have
to break the problem up into bits in order to make some progress.

Here we will focus on the later stages of the radio plasma, in which
the bow shock of the initially overpressurised radio cocoon has
vanished and the cocoon has come into approximate pressure equilibrium
with its surroundings. The radio plasma then becomes Rayleigh-Taylor
unstable and rises under the action of buoyancy forces to form plumes
and mushroom-like structures (Churazov et al. 2001, Br\"uggen \&
Kaiser 2001). The displacement of thermal plasma by the relativistic
plasma will suppress the X-ray emission and create ``holes'' such as
observed in the Perseus cluster (B\"ohringer et al. 1993). Holes in
the X-ray emission are also seen in the radio lobes of the Hydra A
cluster (McNamara et al. 2000).  On the other hand, the hot, buoyant
radio plasma can entrain cooler ambient gas from the centre of the
cooling flow and uplift it to the periphery thus enhancing the X-ray
emission as observed in M87. Recently Saxton, Sutherland and Bicknell
(2001) explained the radio and X-ray features of the middle lobe of
Centaurus A on the basis of a buoyant bubble.

This little studied phase is the phase that we are considering in this
paper. Thus this work differs from the work of Reynolds, Heinz and
Begelman (2001) who simulated an earlier stage in the life of a radio
cocoon, in which the cocoon expands supersonically and causes shocks
in the cluster gas. This phase has also been studied numerically by
Loken et al. (1993), Loken et al. (1995), Clarke, Harris \& Carilli
(1997), Balsara \& Norman (1992), Tregillis, Jones \& Ryu (2001) and
Hardee \& Clarke (1995). In the context of cooling flows jet
simulations have been reported by Soker \& Sarazin (1988), Soker
(1997), Owen \& Eilek(1998), Rizza et al. (2000), Sarazin, Baum, \&
O'Dea (1995), Zhao, Sumi, Burns, \& Duric(1993) and Baum \&
O'Dea(1991). In this paper we will present 3D hydrodynamical
simulations of buoyant gas in the ICM. We extend our previous work
(Churazov et al. 2001, Br\"uggen \& Kaiser 2001) by, first, going to
three dimensions and second by considering a situation in which hot
gas is {\it continuously} injected into a small region close to the
centre of the gravitational potential.

The idea that buoyancy plays an important role in the evolution of the
radio lobes in galaxies has been first proposed by Gull and Northover
(1973) and has been used to estimate the life time of the radio lobes
in M87 by B\"{o}hringer et al. (1995).  Three-dimensional simulations
of hot bubbles in a stratified stmosphere have also been presented by
Brandenburg \& Hazlehurst (2001).

For the simulations presented below a particularly relevant
observational example is perhaps Hydra A -- a radio galaxy associated
with a relatively poor cluster A780. In the radio Hydra A (3C 218) is
a powerful Fanaroff-Riley type I source with extended double radio
lobes (e.g. Ekers \& Simkin 1983, Taylor et al. 1990), suggesting
continuous activity of the central engine for long period of
time. Recent Chandra observations (McNamara et al., 2000, David et
al., 2001) clearly demonstrated the interaction between the radio
emitting plasma and the thermal gas. The morphology of the interacting
region is somewhat different from that observed in M87. In Hydra A
thermal gas has clearly been pushed out of those regions that contain
the radio plumes.

\section{Method}
\label{sec:method}

\subsection{Hydrodynamical simulations}

\begin{figure}[t]

\plotone{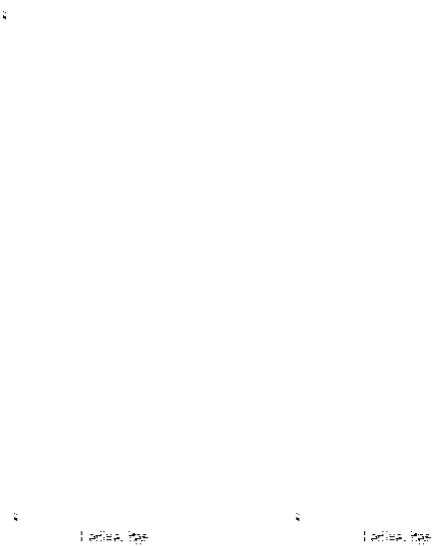}
\caption{Initial distribution of electron density, temperature,
pressure and gravitating mass (adopted from Nulsen and B\"{o}hringer
 1995) assumed in the simulations.
\label{setup}
}
\end{figure}

The simulations were obtained using the parallel version of the
ZEUS-3D code which was developed especially for problems in
astrophysical hydrodynamics (Clarke \& Norman 1994).  The code uses
finite differencing on an Eulerian grid and is fully explicit in
time. It is based on an operator-split scheme with piecewise linear
functions for the fundamental variables. The fluid is advected through
a mesh using the upwind, monotonic interpolation scheme of van
Leer. For a detailed description of the algorithms and their numerical
implementation see Stone \& Norman (1992a, b). In order to study the
spectral ageing of the relativistic gas, the ZEUS code was modified to
follow the motion of `tracer' particles which are advected with the
fluid.\\

In our simulations we employed an ideal gas equation of state and we
ignored the effects of magnetic fields and rotation.  The cooling time
is sufficiently long compared to the time scales considered here
since, for an electron density of 0.02 cm$^{-3}$ and temperature of
the order of 2 keV (see Fig.\ref{setup}) the cooling time is $\sim
5\times 10^8$ years. This is one order of magnitude longer than the
typical duration of the run of $\sim 5\times10^7$ years. We therefore
completely neglected cooling.  The simulations were computed on a
Cartesian grid in three dimensions.  The serial version of the code
was run on SUN ULTRA workstations and the parallel version on an SGI
ORIGIN 3000. The computational domain spanned 30 kpc in height with a
base area of 10 kpc $\times$ 10 kpc. This domain was covered by 450
$\times$ 150 $\times$ 150 grid points. On all boundaries the boundary
conditions were chosen to be outflow conditions.\\

The initial mass and temperature distributions were modelled on the
Virgo Cluster as given by Nulsen \& B\"ohringer (1995) (see
Fig.\ref{setup}). The gas density distribution was then found by
assuming hydrostatic equilibrium to maintain an initially static
model. The pressure scale height varied from around 1 kpc near the
centre to around 40 kpc at the edge of the simulated region. At each
timestep gas was injected into a spherical region with a radius of
$r_{\rm b}=0.7$ kpc at a distance of $d=9$ kpc from the gravitational
center. This injected gas was in pressure equilibrium with its
surroundings. It was hotter than the ambient gas by a factor of
100-400 and had zero initial velocity. At the start of the simulation
24000 tracer particles were uniformly distributed throughout the
computational domain. Subsequently, at certain intervals new,
uniformly distributed tracer particles were added into the spherical
injection region. The gas was treated as a single fluid and was
assumed to obey a polytropic equation of state with $\gamma =5/3$. As
we discuss in detail below, buoyancy drives the hot gas through the
ambient medium as shown in Fig.\ref{density}.\\

Finally, we should address some issues related to the accuracy of
these kinds of finite-difference hydrodynamical simulations.  First,
one can note that the density in the bubble increases.  While the code
can simulate large-scale mixing due to Rayleigh-Taylor and
Kelvin-Helmholtz instabilities, it does not include real diffusion of
particles. Any observed diffusion is therefore entirely numerical. The
boundary between the bubble and the ambient medium also becomes less
sharp as the simulation proceeds due to discretization errors in the
advection scheme. For a test of the advection algorithm in the ZEUS
code see Stone \& Norman (1992).  In simple advection tests it was
found that during the advection of a sharp discontinuity over a grid
of 200 zones the discontinuity is spread over 3-4 grid
cells. Therefore, the small features in our simulation are likely to
be affected by these advection errors whereas the larger features are
not.

Second, numerical viscosity is also responsible for suppressing
small-scale instabilities at the interface between the bubble and the
cooler, surrounding, X-ray emitting gas.  To assess the effects of
numerical viscosity, we have repeated our simulations on grids with
lower resolution. From our experiments we can conclude that 'global
parameters' such as the position and size of the 'mushrooms' as well as
the presence of ``toroidal'' structure are relatively insensitive to
the resolution. The detailed developments of the morphology on small
scales do depend on the resolution and the initial conditions.

\subsection{Simulation of the radio emission}
\label{sec:sync}

Radio synchrotron emission arises from relativistic, magnetised
plasma. The simulations presented in this paper are based on a purely
hydrodynamical scheme with a single, non-relativistic fluid. To obtain
radio maps of the simulations we have to make a number of simplifying
assumptions which we summarise briefly below. These and the method by
which we calculate the synchrotron surface brightness of the simulated
flow are described in greater detail in Churazov et al. (2001).

\begin{itemize}
\item The magnetic field is tangled on scales that are much smaller than the scales of fluid motions. Therefore, the relativistic plasma consisting of magnetic fields and relativistic particles is confined to small `bubbles' intermixed with the non-relativistic, thermal plasma. The size of these bubbles will in practice be set by the tangling scale of the magnetic fields. Thus the two fluids are separated on microscopic scales.
\item The bubbles of relativistic plasma are in pressure equilibrium with the surrounding thermal gas. We further assume that they expand and contract adiabatically with an adiabatic index of $4/3$. For consistency the volume filling factor of the relativistic plasma must be small so that its influence on the total fluid flow is neglegible.
\item The pressure of the relativistic plasma is given by the sum of the energy density of the magnetic field, $u_{\rm mag}$, and that of the relativistic particles, $u_{\rm rel}$. As this is balanced at all times by the thermal pressure, $p_{\rm th}$, we have $3 p_{\rm th} = u_{\rm mag} + u_{\rm rel}$. This sets the initial conditions for the relativistic plasma at the time when it is injected into the computational grid together with the thermal gas. In the following we investigate two cases: Either the magnetic field is uniform throughout the injection region with a value corresponding to the equivalent magnetic field of the Cosmic Microwave Background (CMB), i.e. $B = \sqrt{8 \pi u_{\rm CMB}} = 3.18 (1 +z )^2$ $\mu$G (where $z$ is the cosmological redshift; e.g. Leahy 1991) and so $u_{\rm rel} = 3 p_{\rm th} - u_{\rm CMB}$, or we assume equipartition between the magnetic field and the relativistic particles, i.e. $u_{\rm mag} = u_{\rm rel}= 3/2 p_{\rm th}$.
\item We assume that the relativistic particles are injected into the flow with a power law energy distribution, i.e. $n_e \, d\gamma = n_0 \gamma^{-p} \, d\gamma$ where $\gamma$ is the Lorentz factor of the particles. We set $p=2$ and impose cut-offs so that $1 \le \gamma \le 10^6$. The normalisation of the particle spectrum, $n_0$, is determined by the value of $u_{\rm rel}$.
\end{itemize}

The exact synchrotron spectrum also depends on the energy distribution
of the emitting relativistic electrons. This in turn is determined by
the form of injection of these particles into the flow and subsequent
energy losses due to adiabatic expansion, synchrotron radiation and
inverse Compton scattering of the CMB. To determine the energy
distribution at a given place and time, we use the tracer particles
that are continuously injected with the gas into the flow. These
tracer particles are advected with the fluid and their `pressure
history' is frequently recorded in the course of the simulation. This
completely determines the energy distribution of the relativistic
particles at their respective positions within the flow. Details of
the method can be found in Churazov et al. (2001). Making assumptions
about the strength of the magnetic field, the local synchrotron
emissivity can be calculated from the energy distribution of the
relativistic particles. The radio surface brightness is then
determined by integration along lines of sight through the
computational domain. \\

We assume that the material in the injection region of our simulations
is supplied by the jet of an Active Galactic Nucleaus (AGN). The flow
in the jet from the centre of the source to the injection region is
taken to be ballistic. This implies that the pressure within the jet
declines very rapidly along the flow thus making the jet
underpressured with respect to its environment. The injection region
may then be identified with the location of a strong internal shock
which brings the jet into pressure equilibrium with the surrounding
gas (e.g. Cant{\'o} \& Raga 1991) thus making the jet susceptible to
turbulent disruption. This simplified interpretation is also motivated
by the observations of radio galaxies of type FRI (Fanaroff \& Riley
1974). They often show well-collimated, apparently laminar jets in the
inner regions which start to flare into wider, more turbulent
structures further out (e.g. Muxlow \& Garrington 1991). In most of our
simulations material is injected with zero initial velocity. This
would imply that all bulk kinetic energy is dissipated by the shock at
the location of the injection region.

\section{Results and discussion}
\subsection{Morphology of the plumes}

\begin{figure*}[htp]
\plotfour{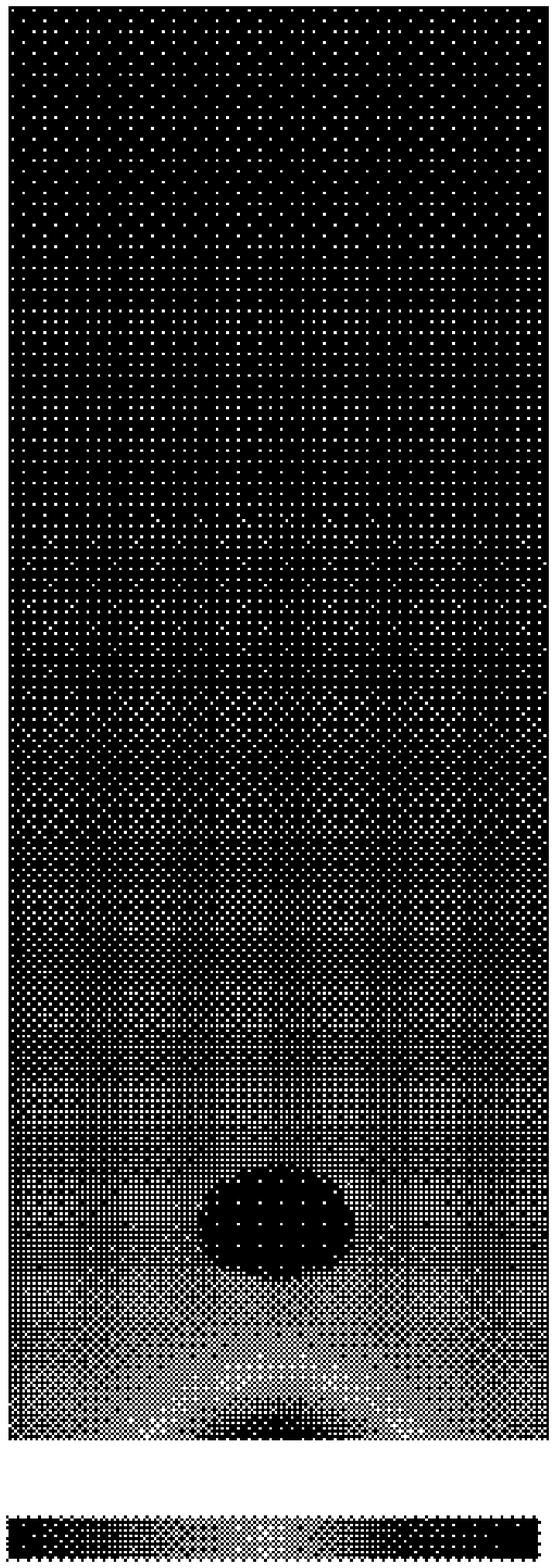}{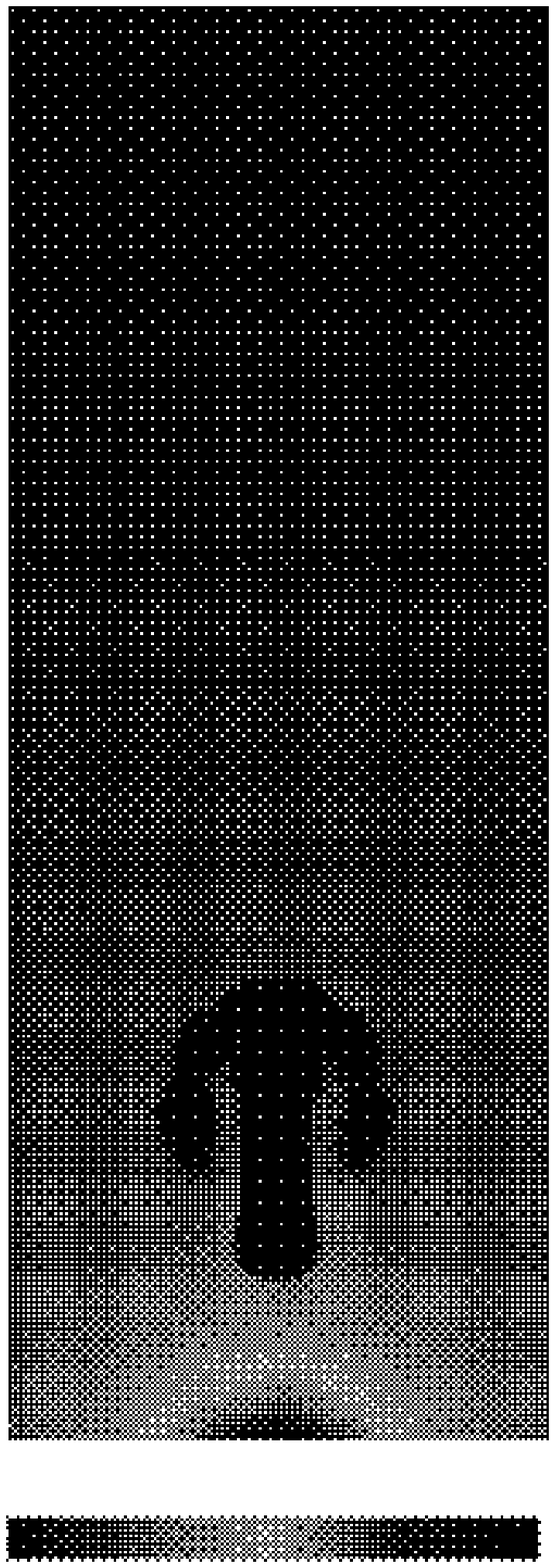}{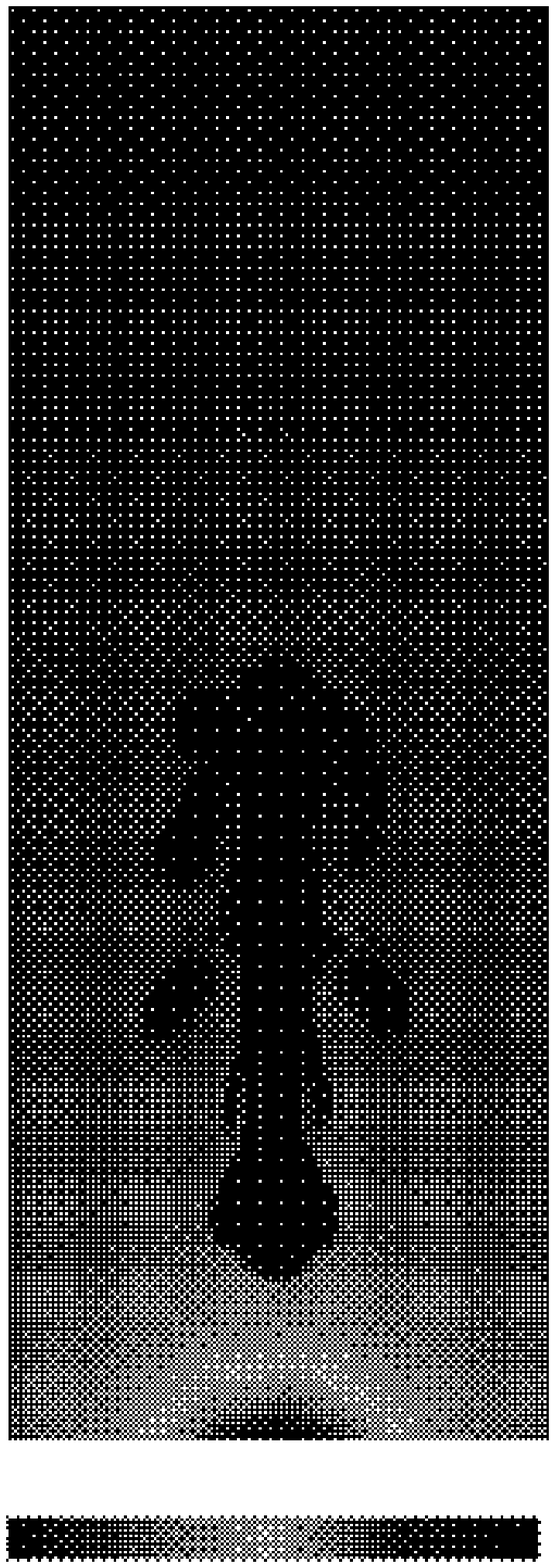}{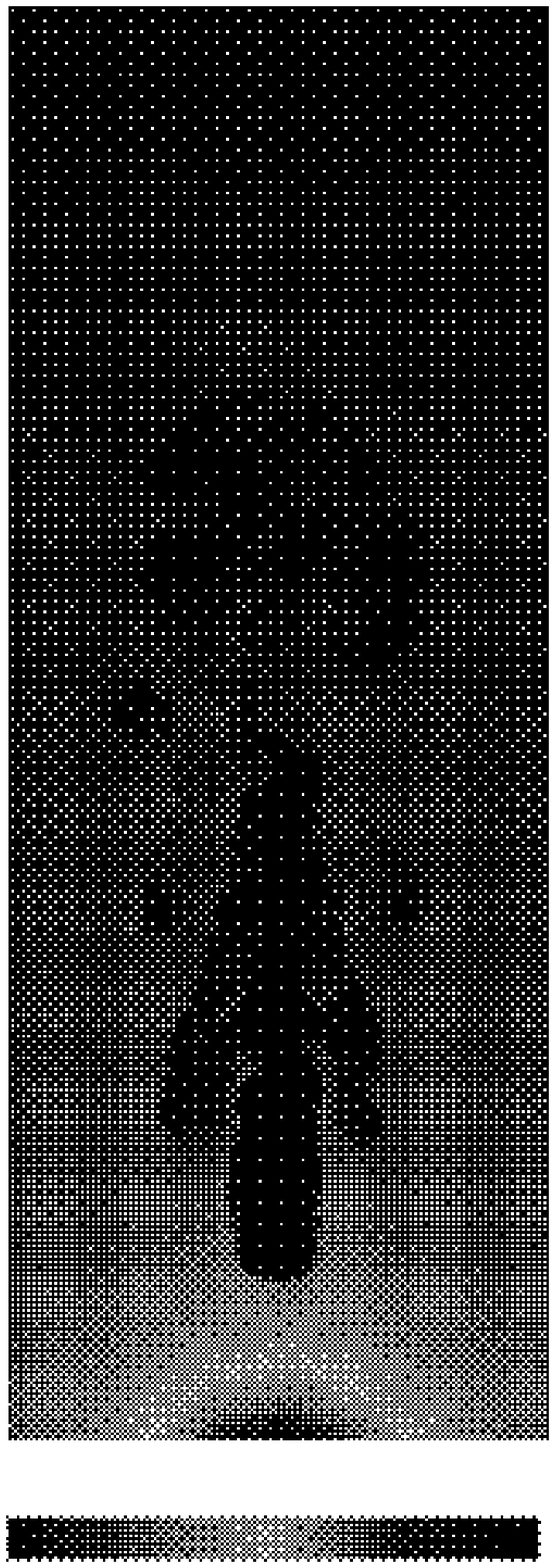}
\caption{Snapshots of density at in a vertical slice through the computational domain for $L=4.4\times 10^{41}$ erg s$^{-1}$ at times 8.36, 25, 41 and 58.5 Myrs.
\label{density}
}
\end{figure*}

\begin{figure*}[htp]
\plotfour{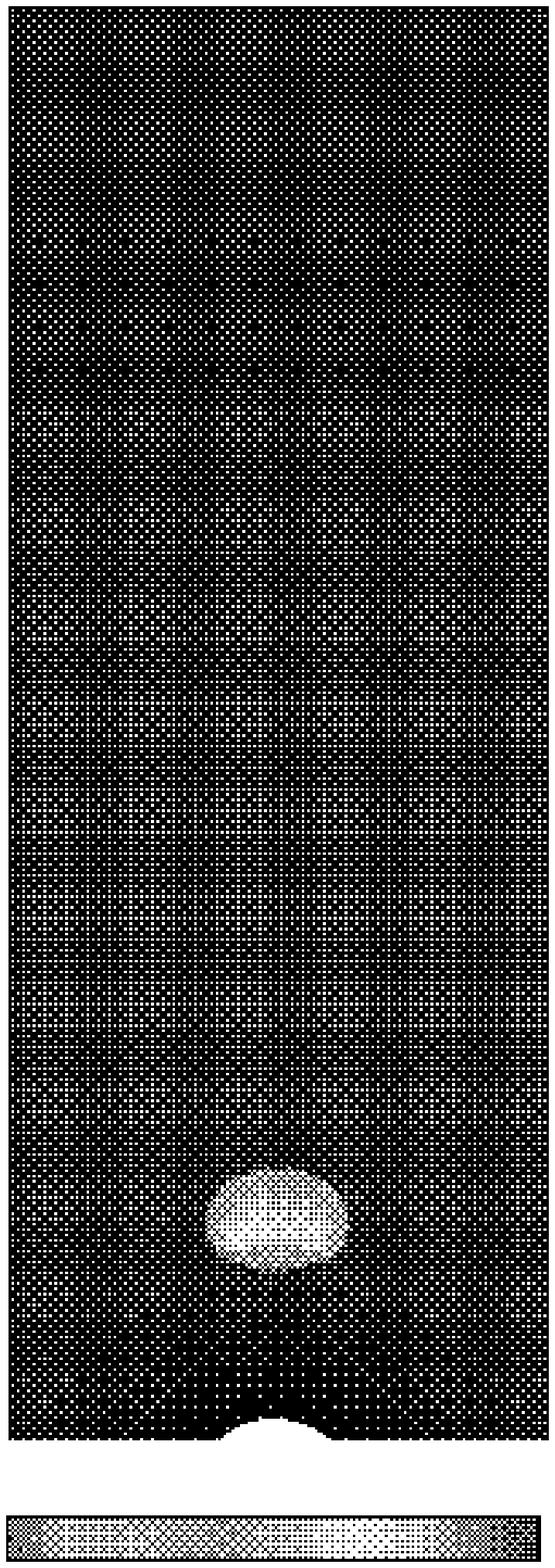}{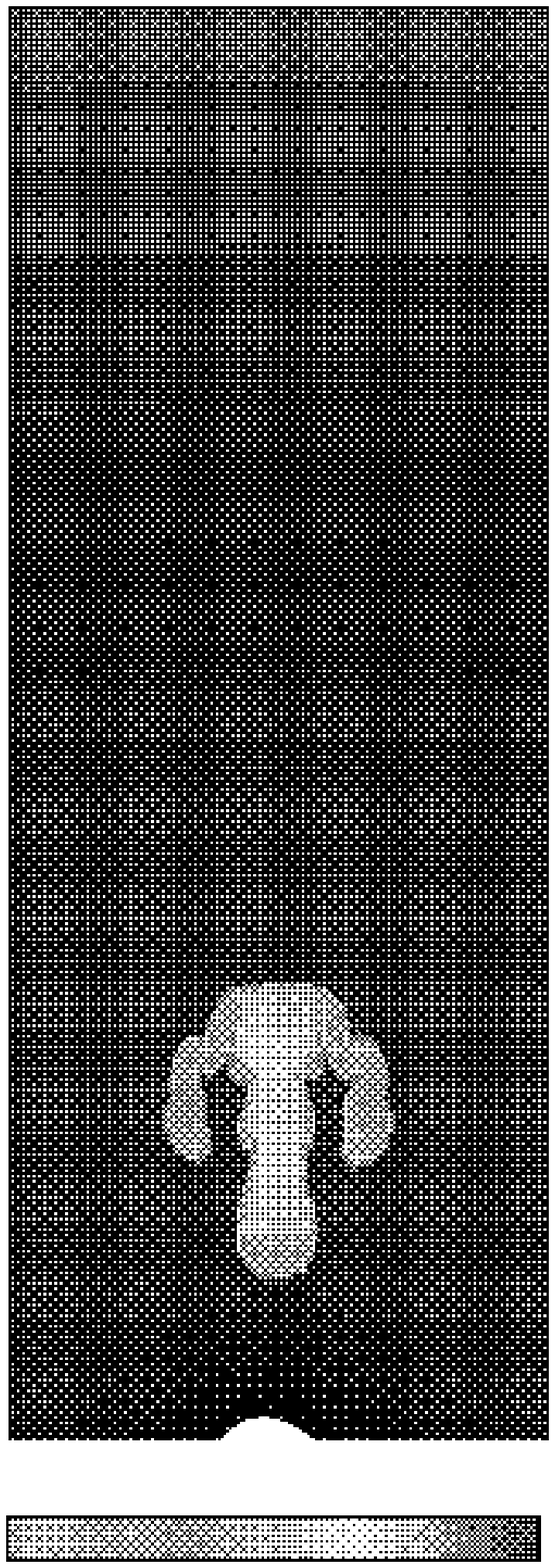}{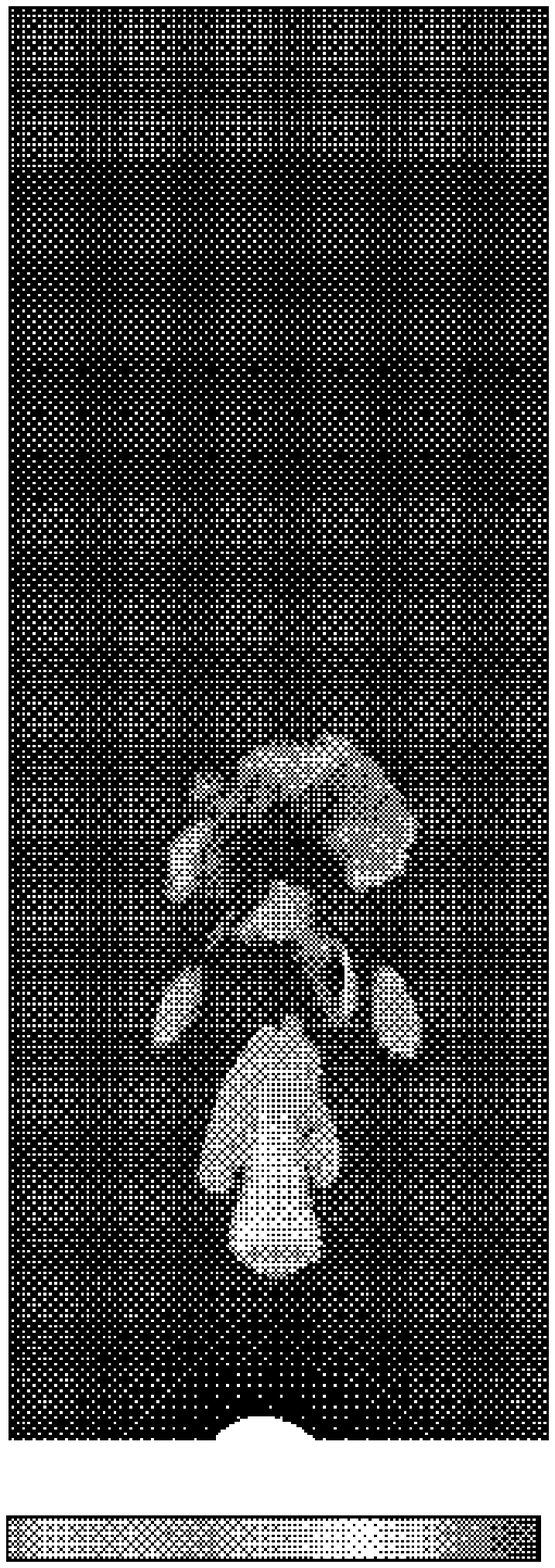}{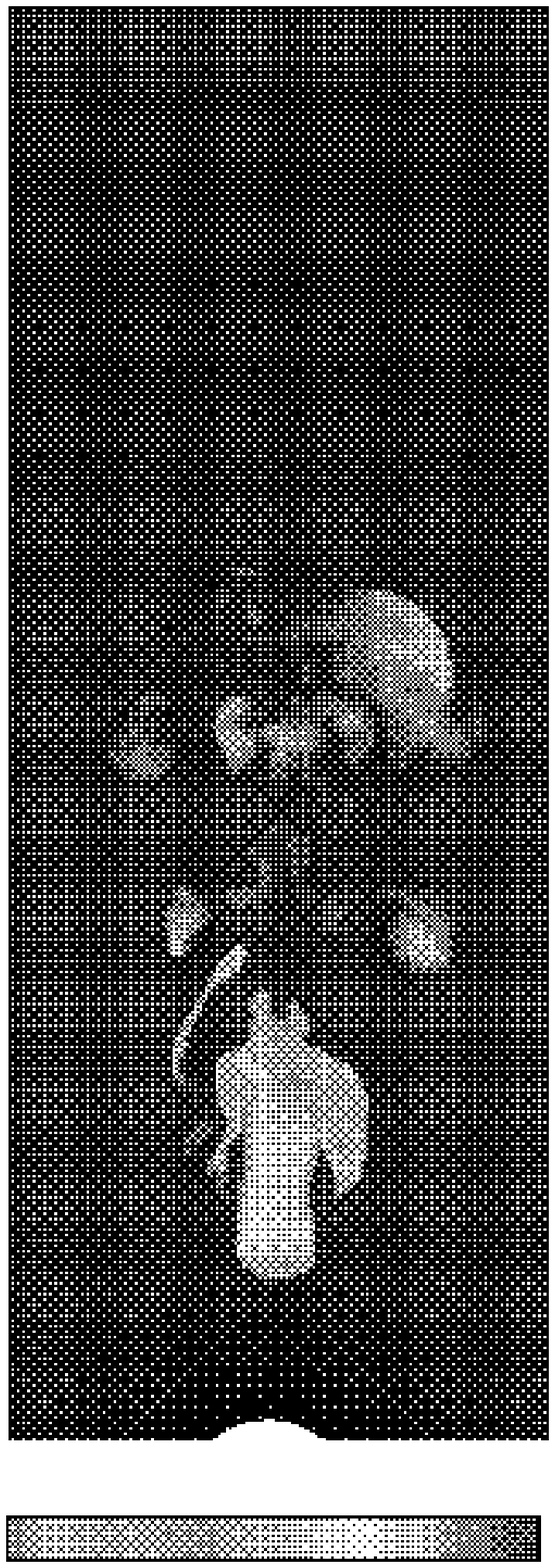}
\caption{Snapshots of specific entropy in a vertical slice through the computational domain for $L=4.4\times 10^{41}$ erg s$^{-1}$ at the same times as in the previous plot.
\label{entropy}
}
\end{figure*}

In Fig. \ref{density} and \ref{entropy} we show snapshots of the
density and specific entropy distribution in a vertical slice through
the computational domain. In this run the luminosity of the injected
gas was $4.4\times 10^{41}$ erg s$^{-1}$ (subsequently called run 1).

One can note that the hot gas can travel to large heights before it is
slowed down by the excess pressure that has built up at the bow of the
plume. The gas is then driven sideways to form a mushroom-like
structure. The plume is also not cylindrically symmetric but meanders
a bit like the smoke from a candle and breaks up into a number of
irregular fragments. \\

The plume or 'nozzle' of hot gas is disrupted by Kelvin-Helmholtz
instabilities as discussed, e.g., by Smith et al. (1983).  The shear
between the rising hot fluid and the ambient medium creates a vortex
boundary layer that destroys the upward flow and mixes the fluid into
the ambient medium. In particular, shear causes the formation of tori
which separate from the main plume. This process, also called ``vortex
shedding'', has been observed and studied extensively (see Norman et
al. 1982). As a rule of thumb one can say that if the plume is thinner
than a scale height of the ambient medium, the jet breaks up into a
turbulent wake of bubbles. Smith et al. (1983) have worked out a
minimum luminosity for a jet to maintain an uninterrupted flow. For a
sound speed ratio between the ICM and the bubble of $c_{\rm b}/c_{\rm
ICM}\sim 10$ as in our simulations, the minimum luminosity is of the
order of $L_{\rm min} \sim L_{\rm c} c_{\rm ICM}/c_{\rm b}$, where $
L_{\rm c} = p_{\rm ICM} c_{\rm ICM} h^2$ with $h$ being a pressure
scale height of the ICM, and $p_{\rm ICM}$ the pressure. (All values
are taken at the location of the bubble.) In our case $h\sim10$ kpc,
$c_{\rm ICM}\sim 4\times 10^7$ cm s$^{-1}$ and $p_{\rm ICM}\sim
10^{-10}$ erg cm$^{-3}$ giving $L_{\rm min} \sim 3.5 \times 10^{41}$
erg s$^{-1}$. So the luminosity in our simulation is near this
analytical estimate for the minimum luminosity for a steady jet. \\

One should also note that the results of the 3D simulations are
different from the results of two-dimensional simulations in which the
boundary coincides with the axis of symmetry. They usually show strong
flows along the boundary(sometimes calles ``axis jets'') and do not
reproduce the correct rise velocities.\\

It is easily shown that the order of magnitude of the terminal
velocity of the gas is
\begin{eqnarray}
v\sim\sqrt{g\frac{V}{S}\frac{2}{C}\frac{\rho_{\rm a}-\rho_{\rm b}}{\rho_{\rm a}}}\sim\sqrt{g\frac{V}{S}\frac{2}{C}}'
\label{eqvr}
\end{eqnarray}
where $V$ is the bubble volume, $S$ is the cross section of the
bubble, $g$ is the gravitational acceleration (we assume the ambient
gas is in the hydrostatic equilibrium), $\rho_{\rm a}$ and $\rho_{\rm
b}$ are the mass densities of the ambient and the bubble gas densities
respectively.  The numerical coefficient $C$ (drag coefficient)
depends on the geometry of the bubble and the Reynolds number. For a
solid sphere moving through an incompressible fluid the drag
coefficient $C$ is of the order 0.4--0.5 for Reynolds numbers in the
range $\sim 10^3$--$10^5$. Here the factor of $(\rho_{\rm a}-\rho_{\rm
b})/\rho_{\rm a}$ can be dropped if the bubble density is low compared
to the ambient gas density. The expression for the terminal velocity
can be rewritten further using the Keplerian velocity at a given
distance from the cluster center: $v\sim \sqrt{(r/R)(8/3C)}v_{\rm K}$,
where $r$ is the bubble radius, $R$ is the distance from the center
and $v_{\rm K}=\sqrt{gR}$ is the Keplerian velocity.  In our
simulations, the typical Keplerian velocity was $\sim$ 400 km/s. Even
though this is just an order of magnitude estimate of the rise
velocity it is approximately correct since the velocity depends only
weakly (as the 0.5 power) on the bubble size and gas density
gradients.\\

In a second simulation the energy is injected into a region of radius
$r=1$ kpc with a luminosity of $3.8\times 10^{42}$ erg s$^{-1}$
(subsequently called run 2). The evolution of the density in this run
is shown in Fig. \ref{MediumE}. Now a much broader plume rises that is
not disrupted and a steady jet forms. Only later secondary
Rayleigh-Taylor instabilities set it that form turbulent mushroom
structures.

\begin{figure*}[htp]
\plotfour{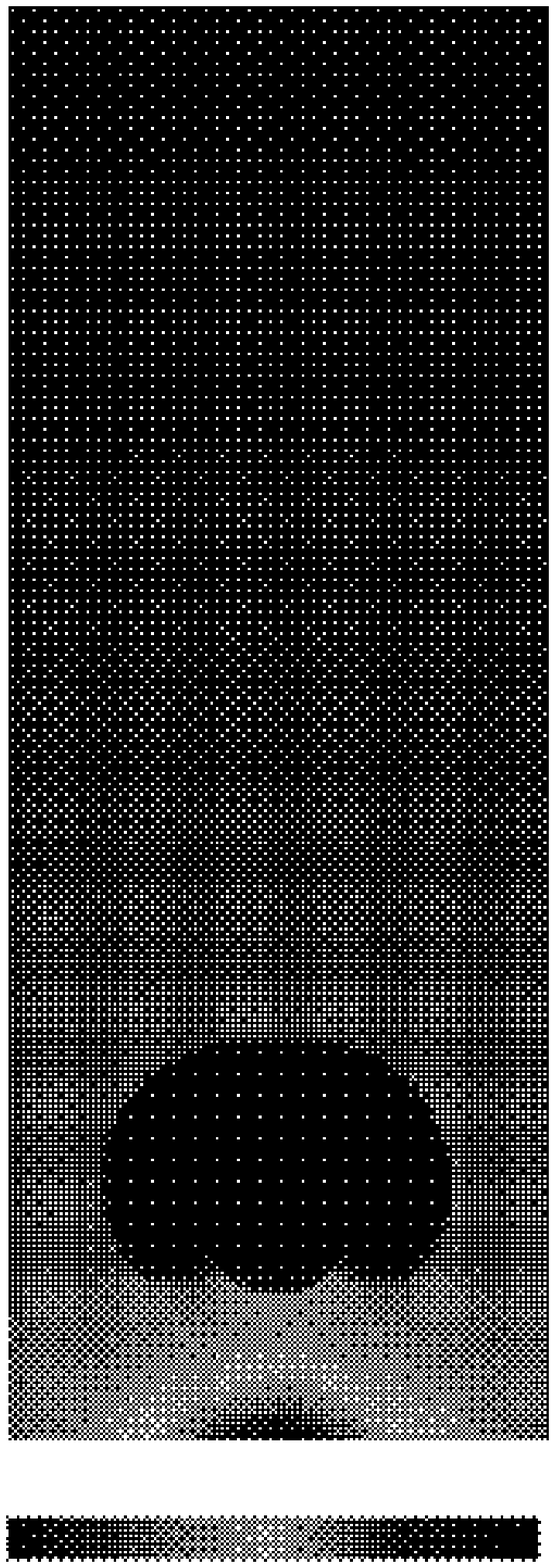}{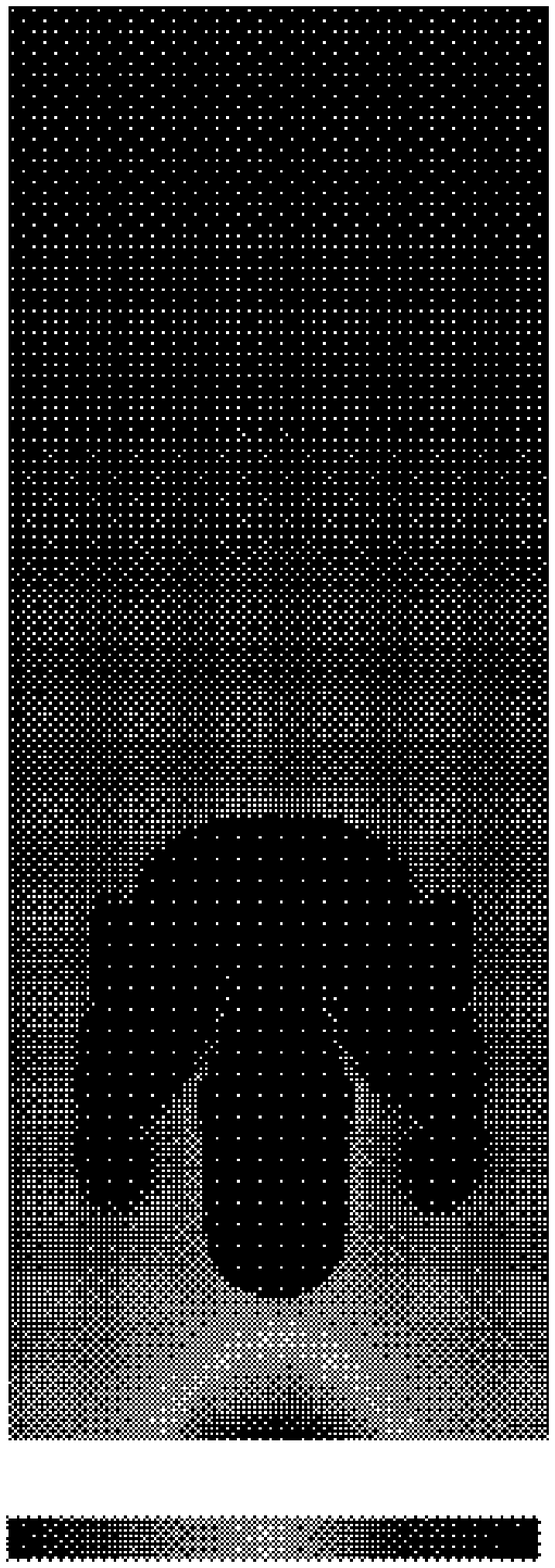}{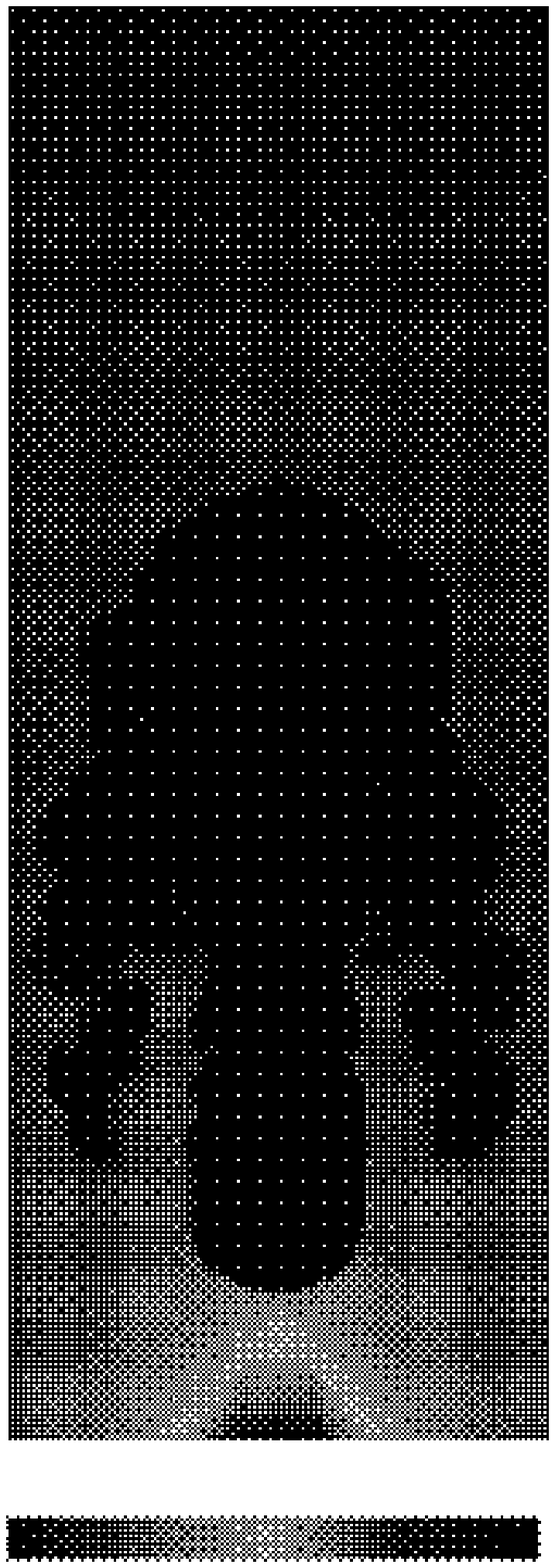}{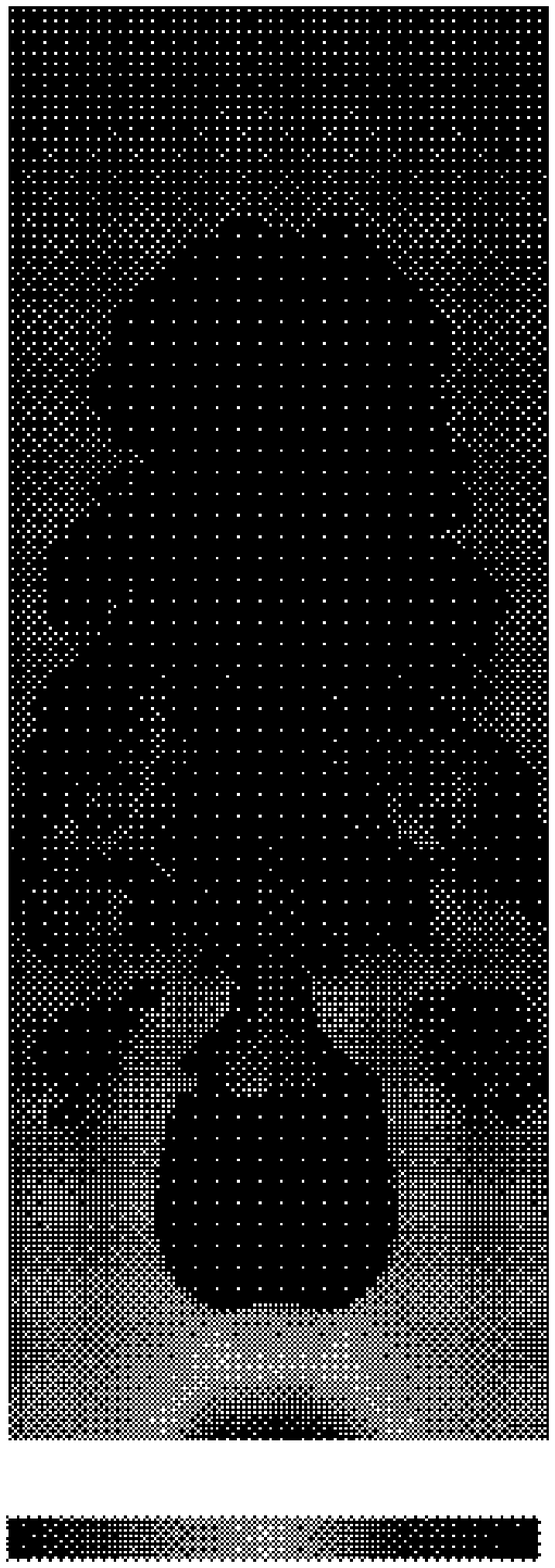}
\caption{Snapshots of the density in a vertical slice through the computational domain for $L=3.8\times 10^{42}$ erg s$^{-1}$ at times 12.5, 25, 37.6 and 50 Myrs.}
\label{MediumE}
\end{figure*}

In this paper we will focus on the two runs described above (run 1 and
2). However, in order to probe the parameter ranges for which our
study applies, we have made two further simulations, in which the
dynamics of the flow turned out to be very different.

\begin{figure}[htp]
\plottwo{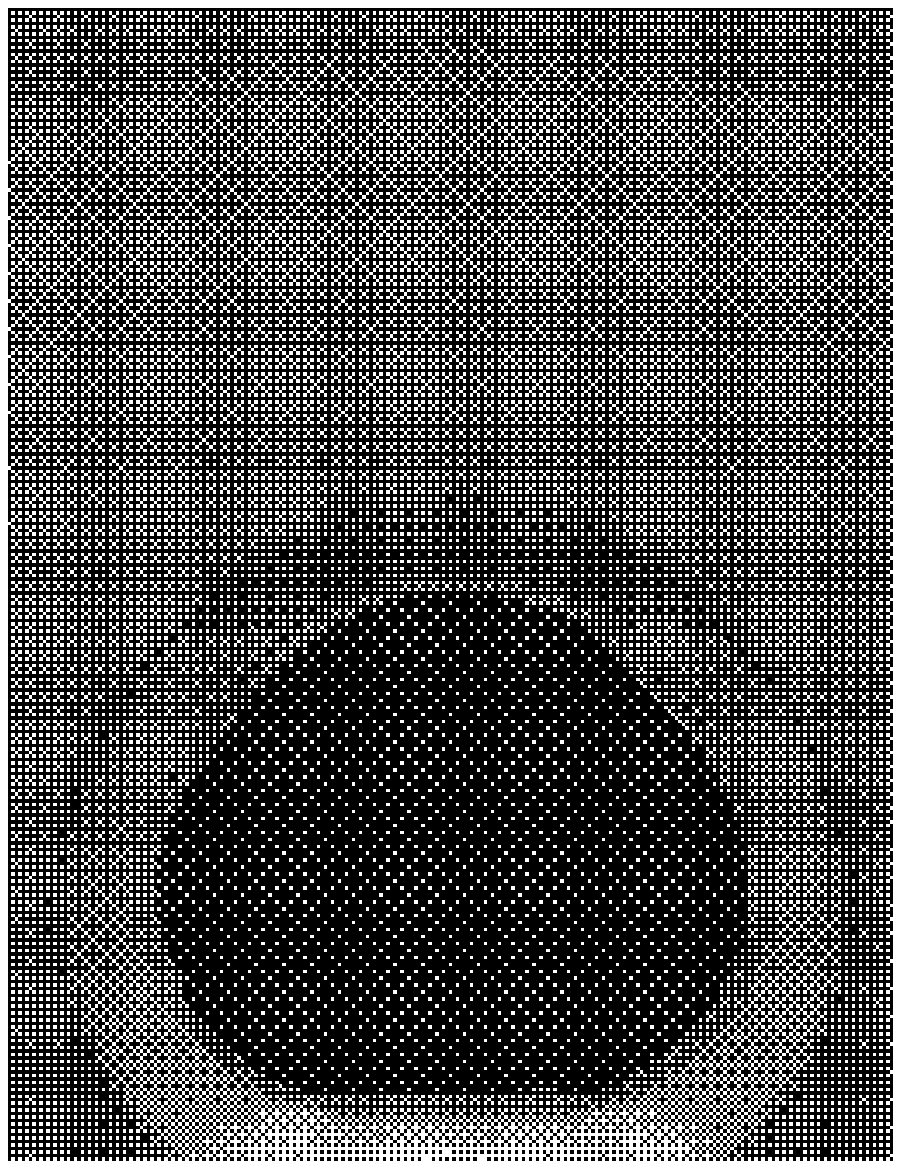}{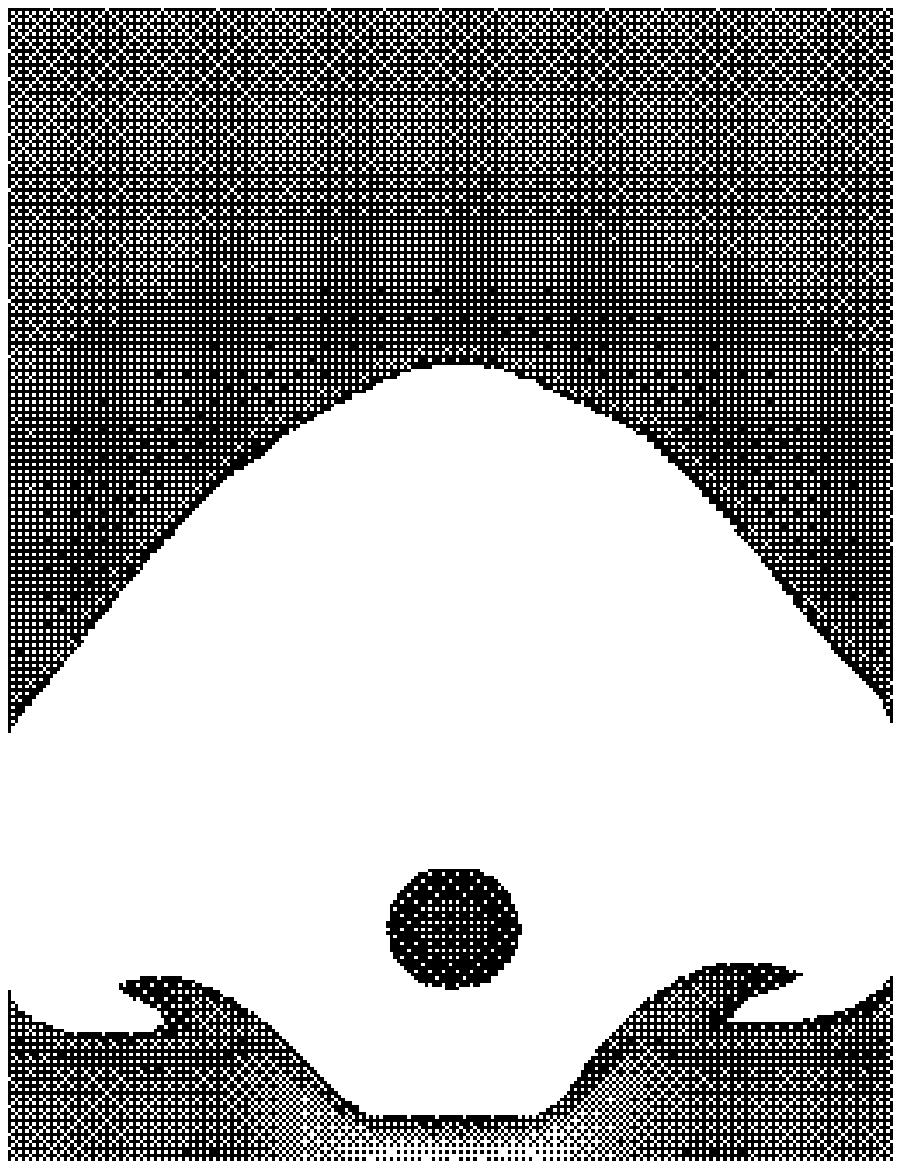}
\caption{Snapshots of the density in a vertical slice through the computational domain for $L=10^{44}$ erg s$^{-1}$ after 5 and 10 Myrs.
\label{HighE}
}
\end{figure}

In Fig. \ref{HighE} we show the density for a run with an even higher
luminosity of $10^{44}$ erg s$^{-1}$ (run 3). In this simulation the
computational domain span 20 kpc in height with a base area of 16 kpc
$\times$ 16 kpc. Now the bubble evolves very differently. A huge
bubble is blown up before it can rise and split into smaller
fractions. This behaviour is in described by Smith et al. (1983) who
estimate that for luminosities $L > 6\ L_{\rm min}$ the injected
matter forms bulky clouds rather than narrow jets. The flow no longer
has the form of a nozzle and is not in equilibrium. The pressure
gradient perpendicular to the symmetry axis cannot confine the jet and
recurrent nozzle choking takes place. These are order-of-magnitude
estimates and we find that the maximum luminosity is higher than the
above estimate by a factor of a few. But our simulations are
in rough agreement with the behaviour predicted by these simple analytical
estimates.

Finally we performed a simulation where the injected material was
given an initial velocity pointing radially outward (run 4). The
velocity was chosen to be three times the sound speed of the local
ambient medium. Thus the kinetic energy of the injected gas was still
many orders of magnitude smaller than the thermal energy. It is
apparent that the jet now evolves on a much shorter time scale. The
flow is much more laminar than in the purely buoyant case and not
governed by Rayleigh-Taylor instabilities. The jet terminates at a
shock that is driven outward with a velocity that is close to the
injection velocity. One can note that the evolution is quite different
in the cases with and without an initial velocity. This is discussed
in more detail in the next section.

\begin{figure*}[htp]
\plotfour{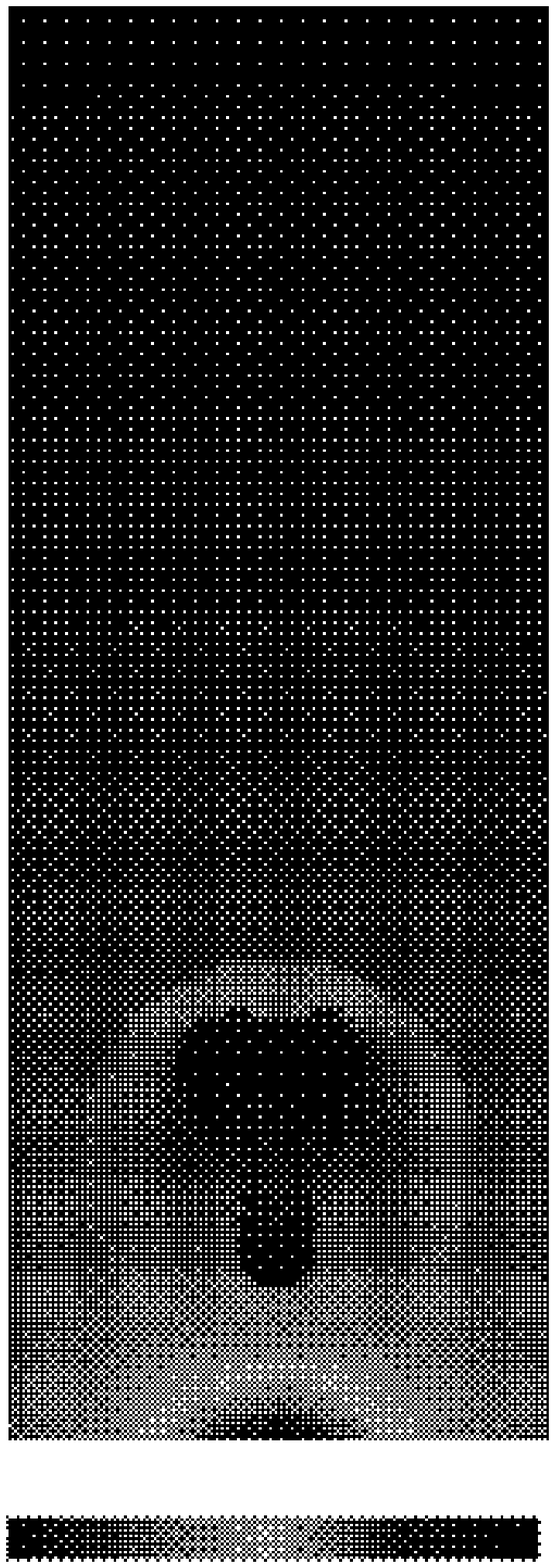}{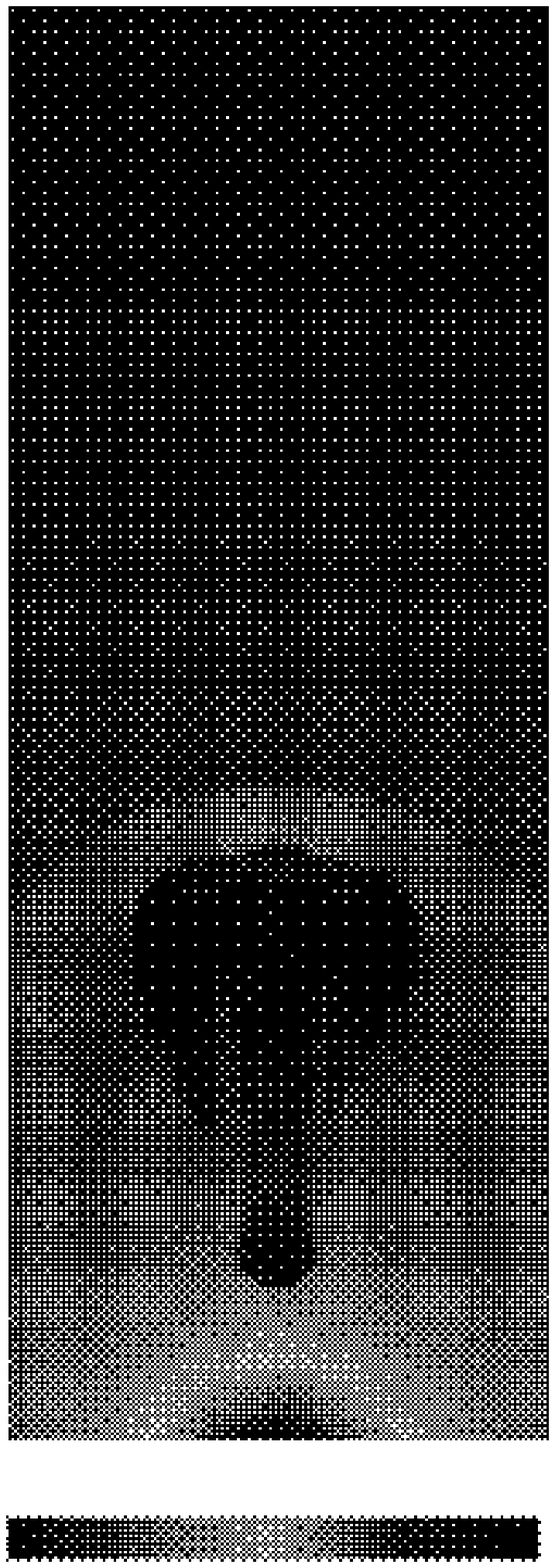}{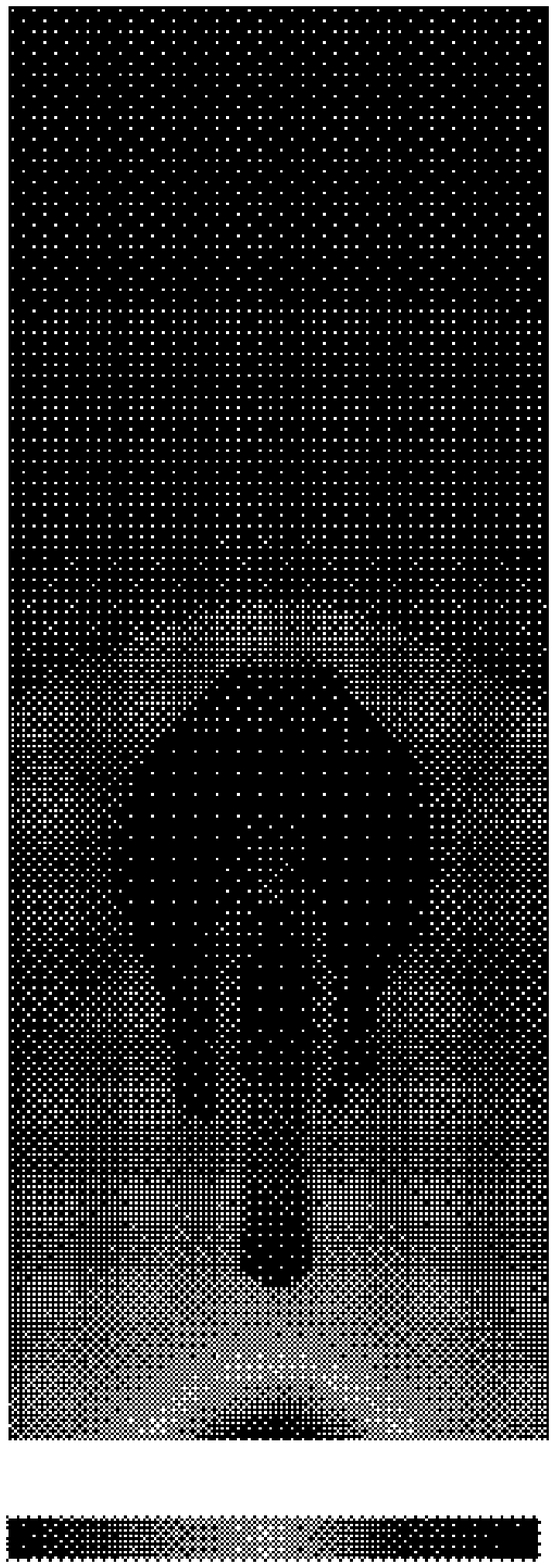}{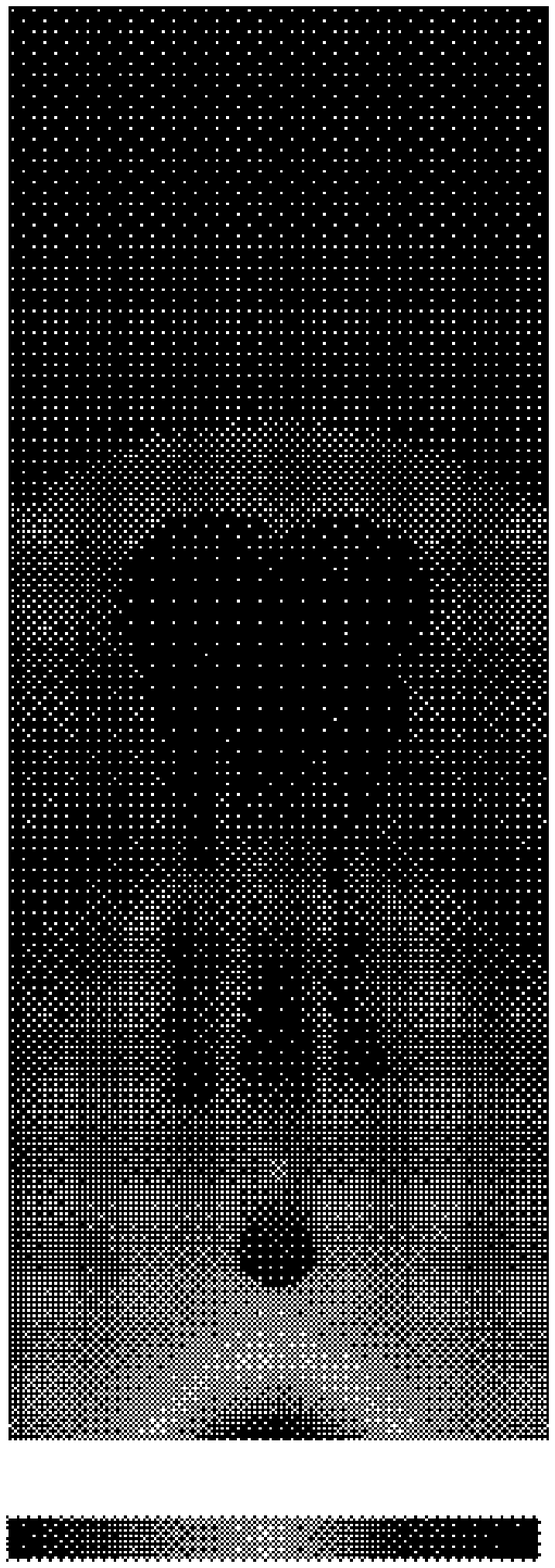}
\caption{Snapshots of the density in a vertical slice through the computational domain for a simulation with an initial velocity of $2.8\times 10^8$ cm s$^{-1}$ and $L=4.4\times 10^{41}$ erg s$^{-1}$ at times 3.8, 7.5, 11.2 and 15.1 Myrs.}
\label{MediumE}
\end{figure*}

\begin{figure}[htp]
\plotone{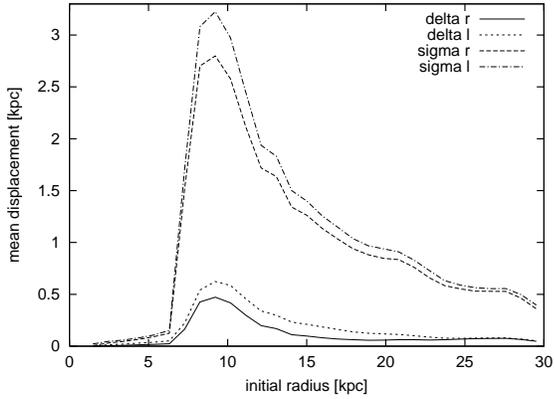}
\caption{\label{fig:tae1} Radial profiles of the average particle
displacement at the end of the simulation time as a function of the
initial cluster radius for run 1. For details and discussion see the text.}
\end{figure}

\begin{figure}[htp]
\plotone{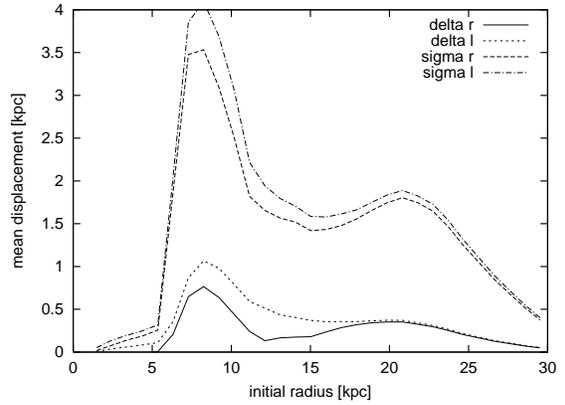}
\caption{\label{fig:tae2} Radial profiles of the average particle
displacement at the end of the simulation time as a function of the
initial cluster radius for run 2.}
\end{figure}

\begin{figure}[htp]
\plotone{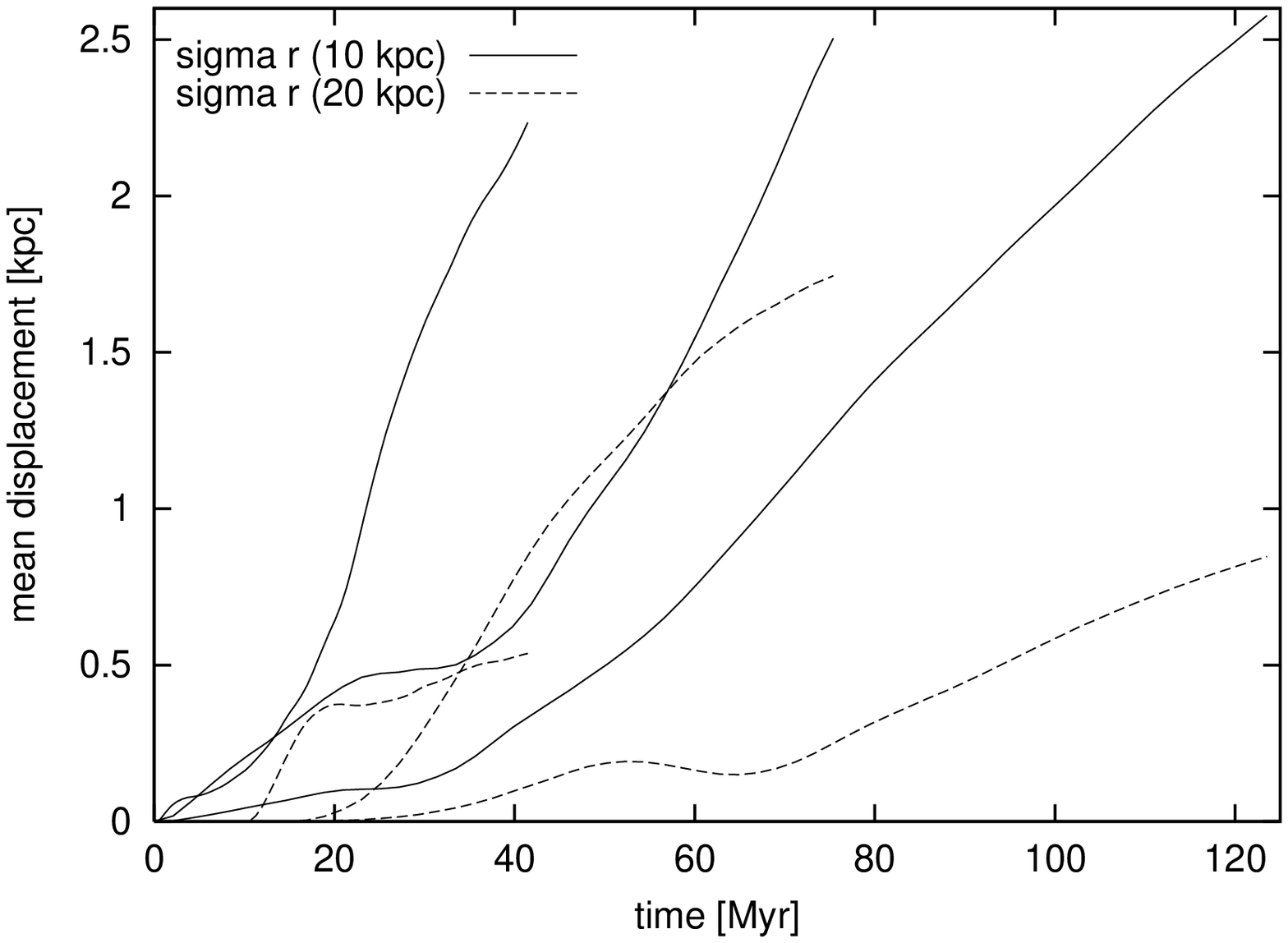}
\caption{\label{fig:tae3} Temporal evolution of the average radial
dispersion of tracer particles for different initial cluster
radii. The lines that terminate at 41 Myrs correspond to run 4, those
that terminate at 75 Myrs correspond to run 2, the other ones to run
1. For details and discussion see the text. }
\end{figure}

The rising bubbles stir convective motions in the cluster atmosphere
and therefore may help to erase the temperature gradient by turbulent
mixing. In order to measure this effect we trace individual fluid
elements in the cluster atmosphere using tracer particles and
calculate the statistics of their displacements. The particles are
binned according to their initial cluster radius $r_{0}$ and then the
mean displacement in radial direction $\delta r (r_{0},t) = \langle
r(r_{0},t)-r_{0} \rangle$ and its dispersion $\sigma_r (r_{0},t) =
\langle (r(r_{0},t)-r_{0})^2 \rangle^{1/2}$ are computed. The brackets
$\langle ... \rangle$ denote the average over all tracer particles
within a spherical half-shell, including virtual particles outside the
simulation volume, which were assumed to be at rest. Furthermore, the
quantities $\delta l (r_{0},t) = \langle
|\vec{r}(\vec{r}_{0},t)-\vec{r}_{0}| \rangle$ and $\sigma_l (r_{0},t)
= \langle |\vec{r}(\vec{r}_{\rm initial},t)-\vec{r}_{0}|^2
\rangle^{1/2}$ were calculated. The profiles of these quantities at
the end of the simulation are shown in Figs. \ref{fig:tae1} and
\ref{fig:tae2}, and their temporal evolution at several initial
cluster radii in Fig. \ref{fig:tae3}.

The relative heights of the profiles in Fig. \ref{fig:tae1} indicate
that there is only a small net displacement of the tracer particles
compared to their dispersion, and that both the dispersion and the net
displacement are predominantly in the vertical direction. One can see
that the dispersion decreases with radius. This is partly because the
estimated volume averages are calculated for spherical shells, whose
volume increases as $r^2$, whereas the simulated jet only affects a
finite and roughly constant area perpendicular to the jet axis.

The temporal evolution of the tracer particle dispersion in radial
direction as shown in Fig. \ref{fig:tae3} shows a time-lag, owing to
the time the radio plasma needs to ascend. Then one can note a small
bump due to the uplifting of gas by the passing bubble. This is
followed by a linear growth of $\sigma_r$. This linear time-dependence
indicates that the regime of turbulent diffusion, in which $\sigma_r
\propto t^{1/2}$, is not yet reached. However, the flow pattern
imprinted by the rising bubbles is able to transport gas to larger
radii. The average radial displacement of the gas within the whole 30
kpc half-sphere is $\bar{\sigma}_r = 0.44 \,$kpc after 75 Myr, and
$\bar{\sigma}_r = 0.91 \,$kpc after 125 Myr at the end of the run
1. At the end of run 2, after 75 Myr, the average displacement is
$\bar{\sigma}_r = 1.38\,$kpc. Since the jet power differed by about
one order of magnitude in the two runs, the mean square displacement
of the environment $\bar{\sigma}^2$ is proportional to the jet
power. Hence the mean displacement varies like the square root of the
power. As the latter is the crucial quantity describing the mixing
efficiency, this suggests that, over a longer period, frequent
low-power activity cycles are more efficient in stirring the
environment than less frequent high-power outbursts of comparable
energy output. We hope that this can be confirmed in future
simulations. 

We have also plotted the evolution of the particles for run 4, where
the radio plasma was given an initial velocity (see
Fig. \ref{fig:tae3}). It is apparent that the efficiency to uplift and
mix gas increases significantly when the plasma carries a small amount
of initial kinetic energy.\\

\subsection{Synchrotron emission from the flow}

Our interpretation of the injection region is that this is the site of
a strong shock caused by a jet emanating from an AGN. This shock may
accelerate electrons and, depending on the composition of the jet,
positrons to relativistic energies. If the jet carries a magnetic
field, we expect to observe radio synchrotron emission from the
injected gas. Using the assumptions and method summarised in Section
\ref{sec:sync} we calculated the surface brightness due to synchrotron
radiation from our simulations. In order to make the results directly
comparable to our earlier simulations of M87 (Churazov et al. 2001),
we assumed an observing frequency of 327 MHz and a distance of 18 Mpc,
appropriate for the Virgo cluster. Here we assume a volume filling
factor of unity for the radio plasma even if we require implicitly a
lower volume filling factor for the plasma to have a negligible
influence on the dynamics. The results can easily be scaled for lower
filling factors.

\begin{figure*}[htp]

\plotwide{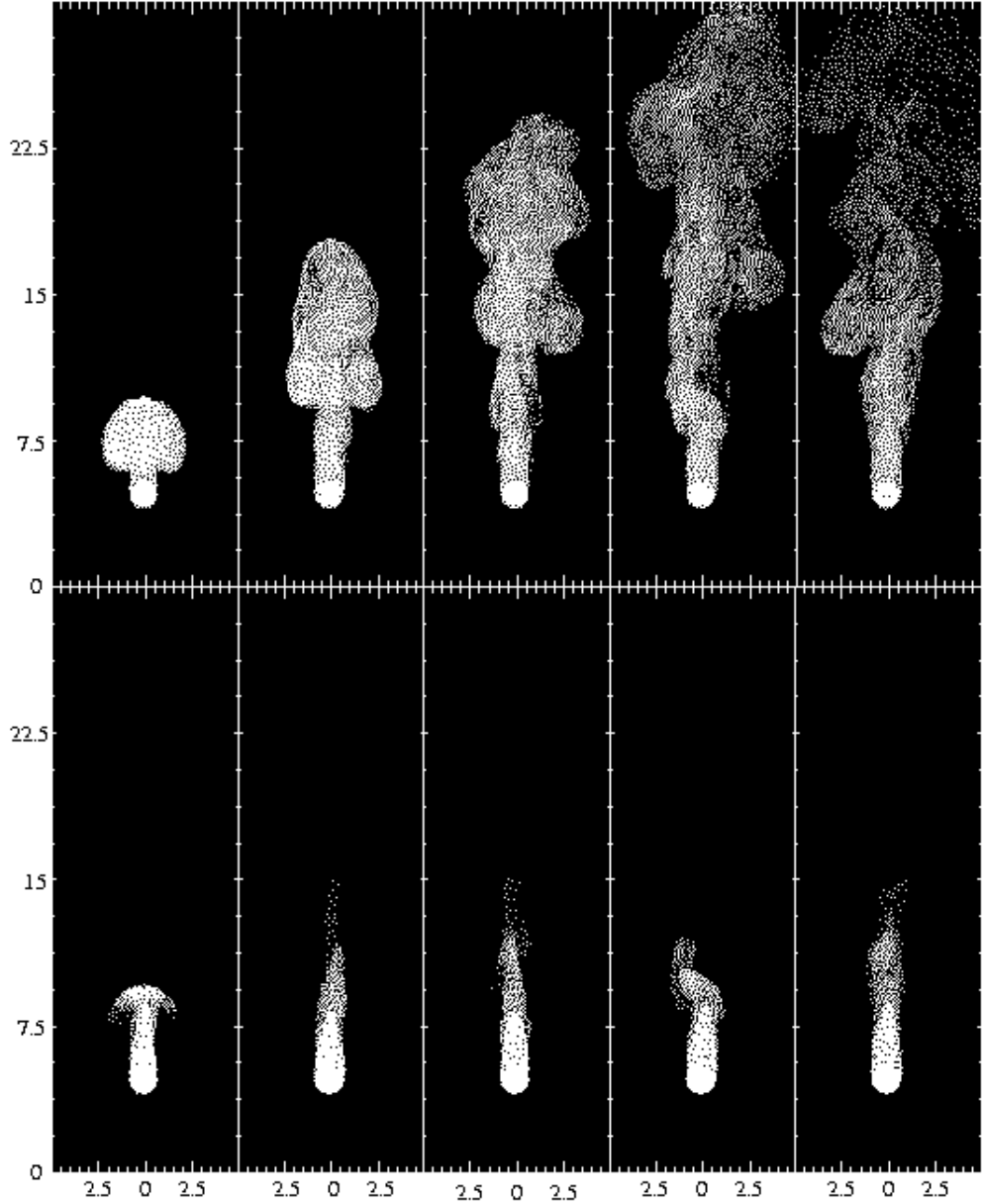}
\caption{Radio surface brightness at times (from left to right) 23.1, 44.1, 65.1, 86.1 and 107.1 Myrs after the start of the simulation (run 1). The axes are labelled in units of kpc. Individual pixels have a size of $0.1\times0.1$ kpc$^2$ which corresponds to $1.15\times1.15$ arcsec$^2$ at the assumed distance of 18 kpc. Upper row: The magnetic field is initially set to the equivalent field of the CMB. Filled contours are logarithmic in steps of 2 starting from 20 $\mu$Jy per pixel (black) to 10.2 mJy (white). Lower row: The energy density of the magnetic field is in equipartition with that of the relativistic particles. Filled contours are logarithmic in steps of 2 starting from 200 $\mu$Jy per pixel (black) to 102 mJy (white).}
\label{stationary}
\end{figure*}

Figure \ref{stationary} shows the results for the radio surface
brightness for our simulation with low luminosity (run 1). In the upper row
the strength of the magnetic field corresponds to the equivalent field
of the CMB. This implies that the field and the relativistic particles
are far from equipartition. The energy stored in the magnetic field
exceeds that in the particles by a factor of several
hundred. Therefore, the plasma is not very efficient in emitting
synchrotron radiation but the lifetime of the relativistic particles
is maximised (e.g. Churazov et al. 2000). For this case, the radio
surface brightness is a good tracer of the pressure distribution
within the flow as in our approach the strength of the magnetic field
is coupled to the thermal pressure. In the last panel `holes' in the
radio structure due to cumulative radiative and expansion losses of
the relativistic particles are clearly visible. Despite these, the
radio structure closely follows the flow structure. In the early
stages the brightest parts of the flow are in the injection region and
at the leading edge of the rising bubble. The former is caused by the
presence of freshly accelerated relativistic particles in a region of
comparatively high pressure. The latter shows the action of the ram
pressure of the external gas being pushed aside in compressing the
buoyant bubble. Fluid instabilities cause the flow to develop
turbulent structures at later times and these are traced by the radio
emission. Local pressure enhancements caused by the turbulent motion
are visible as slightly brighter filaments in the radio maps.

In the lower row of Figure \ref{stationary} we show the results for
the same simulation but now we assume initial equipartition
between the energy in the magnetic field and the energy of the relativistic
particles. This is close to the energy distribution which is most
efficient in producing the synchrotron radiation (e.g. Longair
1994). Again the injection region is the most prominent feature in the
radio maps. However, radiative energy losses are now much more
severe. Thus even in the early stages of the simulation significant
parts of the buoyant flow are not observable in the radio. Only until
about 40 Myrs into the simulation can the ram-pressure compressed
leading edge of the bubble be recognised in synchrotron emission. At
this time it has reached a height of roughly 19 kpc. At later times
the energy losses of the relativistic particles are too strong and we
only see the lower parts of the `chimney' excavated by the buoyant
material. However, the twisting of the flow in the cimney induced by
fluid instabilities is clearly visible. The radio emission gives the
impression of a candle flame under the influence of some air draft
passing by.

\subsection{Comparison with observations of radio galaxies}

\begin{figure*}[htp]

\plotwide{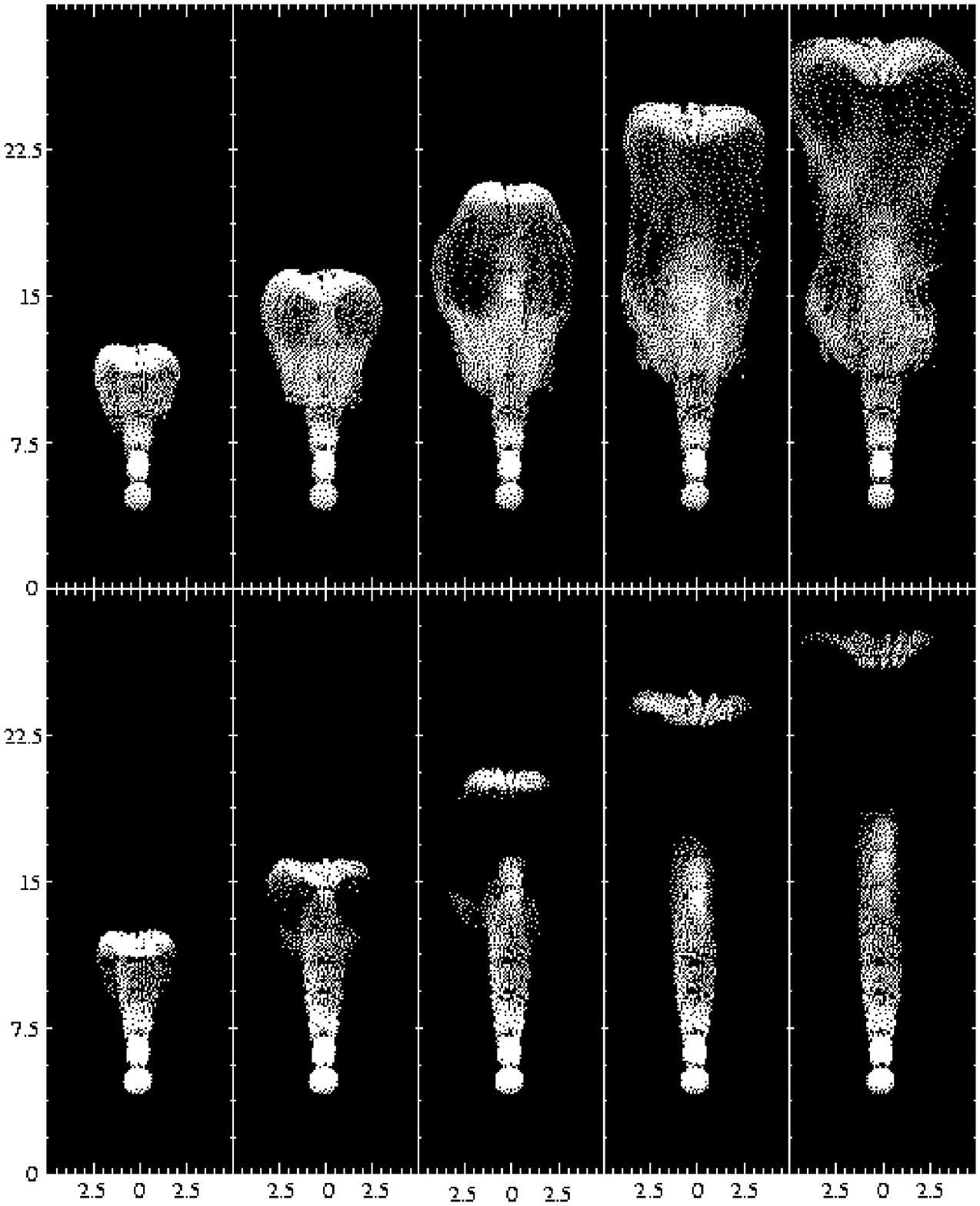}
\caption{Radio surface brightness maps for the low-resolution simulation with non-zero initial velocity of the injected gas (see text). Contours and labels as in Figure \ref{stationary}. Simulation time of panels from left to right: 6.3, 11.3, 16.4, 21.4 and 26.5 Myrs.}
\label{velocity}

\end{figure*}

The radio maps in the upper row of Figure \ref{stationary} resemble
the morphology of radio galaxies of type FRI. The gas flow is
relatively laminar after leaving the injection region but becomes
increasingly turbulent in the outer regions. However, there are a
number of problems with a direct comparison of the simulations with
observations of FRI objects. Firstly, our simulations assume a
vanishing initial flow velocity in the injection region. This is
certainly not the case in FRI objects where the flow velocities even
after the jet flaring are at least mildly relativistic. All
outward motion in our simulations is induced by buoyancy of the hot
gas in the injection region. It is therefore not surprising that the
rise of the turbulent flow region out to tens of kiloparsecs proceeds
on timescales exceeding the lifetime of the relativistic electrons
responsible for the synchrotron emission. Unless there is significant
re-acceleration of relativistic particles in the turbulent part of the
flow, models with purely buoyant motion will not be able to explain
observations of FRI objects spanning more than 1 Mpc (e.g. 3C 31,
Leahy, DRAGN Atlas\footnote{Available at
http://www.jb.man.ac.uk/atlas/}). Moreover, the energy injection rates
in our simulations are much smaller than the typical energies of radio
jets of about $5\times 10^{44}$ erg s$^{-1}$ (Owen, Eilek \& Kassi
2000). However, what these simulations tell us, is that it is possible
to inject a considerable amount of energy into clusters without
producing strong observational signatures. It is thus possible that a
significant amount of energy may be hidden in clusters which is not
easy to detect (also see En{\ss}lin 1999).\\

To investigate the effect of a non-zero initial velocity we performed
a simulation in which the material in the injection region is given an
initial velocity that points radially outward (upward in all Figures
presented in this paper). Figure \ref{velocity} shows the
synchrotron radio maps for this simulation (run 4). The brightest feature is
the strong compression in the flow just above the injection
region. There is a secondary, weaker compression within the developing
laminar flow that moves outwards and produces another enhancement of
the surface brightness.  From theoretical arguments it is well known
that the radius of a collimated, laminar gas flow oscillates about its
equilibrium radius, i.e. the radius which would allow the flow to be
in pressure equilibrium with its surroundings (e.g. Bridle, Chan \&
Henriksen 1981). The flow expands and accelerates until it is
underpressured with respect to its surroundings. A compression and
deceleration region will follow resulting in the pressure inside the
flow to rise above that of the ambient medium. This behaviour can
recur many times.  In our simulations it leads to the formation of
bright `knots' within the laminar gas flow. In supersonic flows this
expansion and re-collimation often leads to the formation of shocks
(e.g. Wilson \& Falle 1985). Although the injected material in our
simulation reaches Mach numbers of up to 5 in the acceleration
regions, the re-collimation curvature radius is so small that the flow
does not develop a shock (Icke 1991).

The leading edge of the rising bubble proceeds at a much higher
speed than in the purely buoyant case. The rise of the flow ends in a
shock which is clearly visible due to the strong compression of the
gas it causes. Unlike in the purely buoyant case, the growth time of
the fluid instabilities is now too long to significantly disrupt the laminar
flow from the injection region. At later times the gas flow is
increasingly protected by the developing lobe of material flowing back
from the region where the rising gas impacts on the surrounding
medium. These features, a shock at the end of the laminar flow and the
lobe inflated around the rising flow, are well-known features of radio
galaxies of type FRII. They have been studied in great detail in
numerous numerical simulations. However, it is interesting to note
that the transition from turbulent flow to a self-protecting and
self-collimating laminar one occurs quite abruptly when adding an
initial velocity to the injected gas. Since the gas in the jets of
FRI-type radio galaxies is known to have high velocities, they most
probably become turbulent solely through the destabilizing entrainment
of gas from their surroundings.

\subsection{The rising bubble in X--ray emission}

\begin{figure}[t]

\plottwo{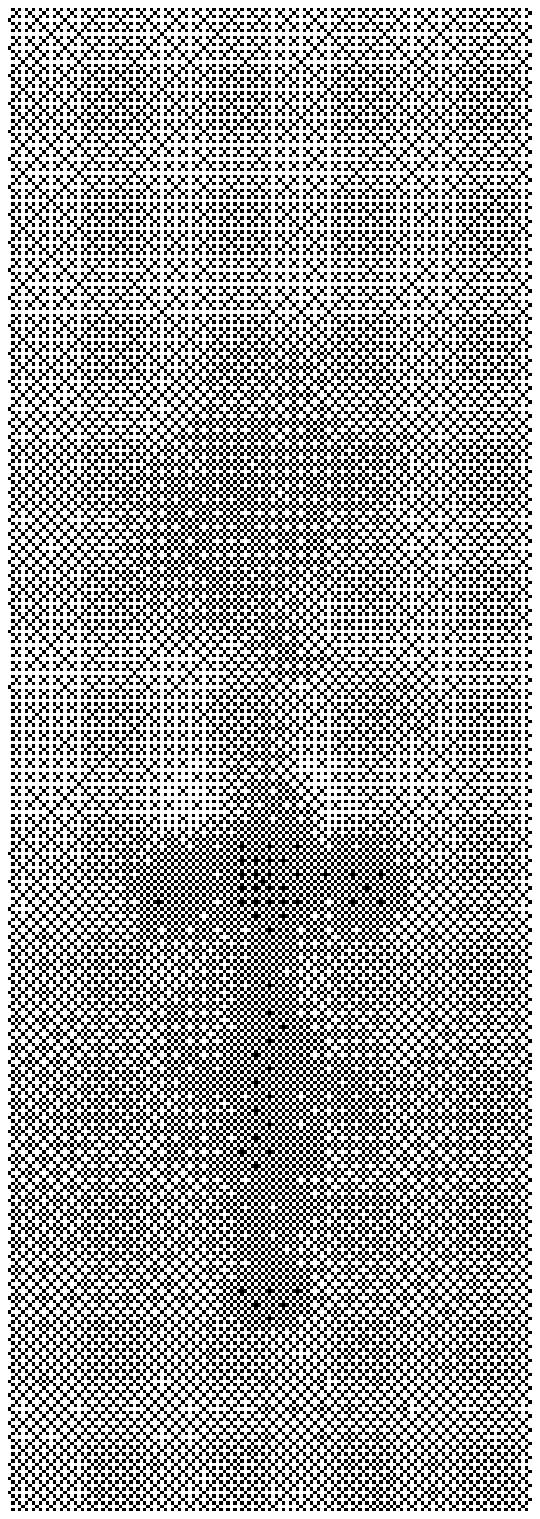}{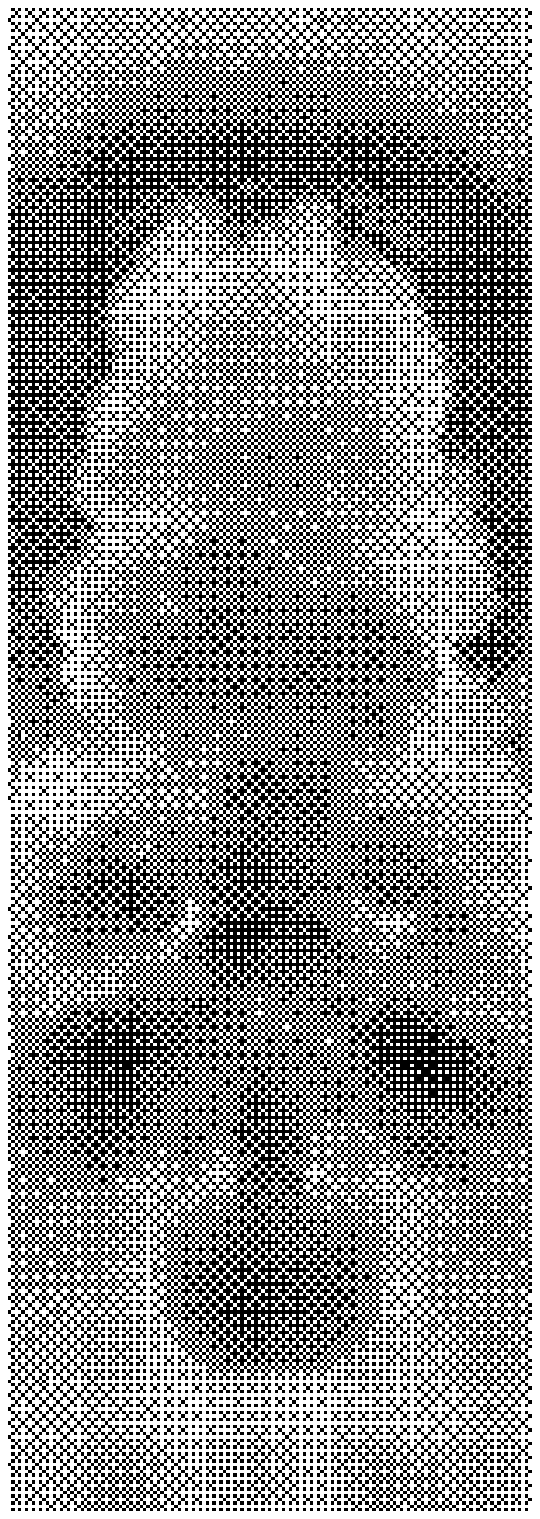}
\caption{X-ray surface brightness relative to the undisturbed cluster for run 1 (left) and run 2 (right).
\label{Xray1}
}
\end{figure}

In Fig.\ref{Xray1} we have plotted the X-ray surface brightness
associated with our simulation 1 and 2.  The surface brightness was
calculated by integrating the emissivity along the line of sight. The
density and the temperature outside the computational volume were
assumed to take their undisturbed values. Red/blue colour correspond
to regions where the X-ray surface brightness is 20-25 \%
brighter/dimmer than the undisturbed brightness. Green colour
indicates that the surface brightness is close to the undisturbed
value. One can note that the effect of the hot plasma on the X-ray
surface brightness is relatively small. The luminosities that we have
considered here are 1 - 2 magnitudes smaller than that of a typical
radio jet or the X-ray luminosity. Thus, we can conclude that radio
plasma injected with luminosities $\sim 10^{42}$ erg s$^{-1}$ is not
easy to detect in X-ray observations.

\section{Conclusions}

In this paper we presented high-resolution 3D simulations of buoyant
bubbles in a typical cluster environment. The aim was to compute the
appearance of these buoyant bubbles in radio and x-ray observations,
and to study the impact of the bubbles on the ICM. Depending on the
luminosity of the injection region, we found a qualitatively different
behaviour of the fluid flow. For low luminosities of around $10^{41}$
erg s$^{-1}$ the thin stream of hot gas breaks up into a series of rising
bubbles. For higher luminosities a broad, steady stream develops that
forms a mushroom-like structure via Rayleigh-Taylor instabilities. For
luminosities greater than $\sim 10^{43}$ erg s$^{-1}$ huge cavities
are blown up before they can rise through the medium. These findings
are in rough agreements with estimates by Smith et al. (1983).\\

We produced radio maps of the rising bubbles assuming that
relativistic particles and magnetic fields are injected alongside the
thermal gas in the injection region. Following the energy loss history
of these particles, we could calculate the radio emission. For very
low magnetic fields (comparable to the equivalent field of the CMB) we
find that the radio emission closely traces the pressure distribution
within the flow. Filaments of enhanced radio emission can be
identified as turbulent compression regions. For equipartition, the
life time of the relativistic particles is significantly shorter than
the flow timescale and only those parts of the flow close to the
injection region are detectable.

The morphology of the radio emission resulting from our simulations is
reminiscent of FRI-type radio galaxies. However, all fluid motion is
induced by buoyancy alone and the involved timescales are thus so long
that it is unlikely that any such source would be detectable in the
radio. If the gas is injected with a non-zero outward velocity, then
the flow timescale reduces greatly and may become shorter than the
lifetime of the relativistic particles. However, in this case the flow
shows the typical self-collimation through alternating expansion and
contraction zones. The morphology of such a flow is more similar to
that of FRII-type radio galaxies. 

In this paper we have not addressed the possibility of re-acceleration
of relativistic particles within the turbulent flow. This effect may
make the flow detectable in the radio even after the initially
accelerated relativistic particles would have faded. However, these
processes are not well understood and we therefore do not attempt to
include them here.\\

In addition to the radio maps, we produced maps of the X-ray surface
brightness. Similar to the radio maps, we found that our bubbles do
not leave a strong signature in the surface brightness.  What this
tells us is that it is possible to hide a lot of energy in form of
buoyant bubbles in the ICM without producing noticable observational
features in the radio and X-ray band. However, for some clusters mass
estimates have been derived from gravitational lensing. They were
found to be in quite good agreement with mass estimates from X-ray
observations suggesting that bubbles do not provide a significant
fraction of the pressure support in the cluster. Nevertheless, it is
an important point that in clusters a lot of energy in the form of hot
bubbles may remain unnoticed.\\

We computed the average diplacement and dispersion of the ambient gas
by the bubbles using tracer particles. The buoyant plasma was shown to
be able to transport ambient gas to larger radii. Comparisons of the
transport efficiencies of different runs suggests that, over a longer
period, frequent low-power activity cycles are more efficient in
stirring the environment than less frequent high-power outbursts of
comparable energy output. Furthermore, we found that the transport is
most efficient for gas near the injection point of the gas. Therefore,
in a cooling flow cluster with a central galaxy, the most rapidly
cooling gas is also most efficiently displaced.\\

It is difficult to estimate accurately what fraction of the energy of
these subsonic motions will eventually be dissipated locally and what
fraction will be carried (e.g. by sound waves and large scale motions)
away from the cooling flow region. If the jet power is indeed $\sim
10^{44}$ erg/s then even a modest 10\% efficiency of local dissipation
into heat should be enough to exceed the radiative cooling of the gas.

However, we find that, with the given numerical scheme, it is
difficult to estimate the dissipation reliably. The numerical
diffusivity and viscosity impose uncertainties that are difficult to
quantify exactly. We therefore urge great caution towards energy
dissipation rates that are inferred from these kinds of
simulations. In order to answer the important question of the energy
dissipation into the ICM one needs simulations with a higher
accuracy and higher resolution. This is work in progress and will be
reported at a later time.

\acknowledgments 

Some of the computations reported here were performed using the UK
Astrophysical Fluids Facility (UKAFF). This work was supported by the
European Community Research and Training Network `The Physics of the
Intergalactic Medium'. MB also acknowledges support through a Junior
Research Fellowship from Churchill College and thanks Jim Pringle for
helpful discussions.

\label{lastpage}

\end{document}